\documentclass[reprint,aps,prd,onecolumn,notitlepage,showpacs,nofootinbib,preprintnumbers,superscriptaddress]{revtex4-1}
\usepackage{amsmath}
\usepackage{amsfonts}
\usepackage{amssymb}
\usepackage{latexsym}
\usepackage{color,xcolor}
\usepackage{graphicx}
\usepackage{bm}
\usepackage{epsfig}
\usepackage[english]{babel}
\usepackage{hyperref}
\usepackage{times}
\usepackage{comment}

\def\O{\mathcal{O}}
\def\D{\mathcal{D}}
\def\S{\mathcal{S}}
\def\R{\mathcal{R}}
\def\GB{\mathrm{GB}}
\def\lagb{\lambda_{\GB}}
\def\gagb{\gamma_{\GB}}
\def\Nsh{N_{\sharp}}
\def\ptr{(\pi TR)}
\def\pt{\pi T}
\def\fw{\mathfrak{w}}
\def\fq{\mathfrak{q}}
\def\fz{\mathfrak{z}}
\def\tom{\tilde{\omega}}
\def\tq{\tilde{q}}
\def\tu{\tilde{u}}

\def\sqrtz{\sqrt{1-4\lagb+4\lagb\fz^{-4}}}
\def\lncom{\ln\frac{\left(1+\fz^{-2}\right)\left(\sqrtz+\fz^{-2}\right)}{\left(1-\fz^{-2}\right)\left(\sqrtz-\fz^{-2}\right)}}

\begin{document}
%\preprint{\vbox{ \hbox{}   \hbox{} }}

\title{Shear quasinormal modes of Gauss-Bonnet black brane: the first post-hydrodynamic order}
%\author{Yu-Chen Ding}
%\author{Li Li}
\author{Towe Wang}
\email[Electronic address: ]{twang@phy.ecnu.edu.cn}
\affiliation{Department of Physics, East China Normal University,\\
Shanghai 200241, China\\ \vspace{0.2cm}}
\date{\today\\ \vspace{1cm}}
\begin{abstract}
Assuming $\omega=\sum_n C^{(n)}q^{2n}$ in the low-frequency limit, we apply the refined recipe to compute the dispersion relation of shear quasinormal modes of the Gauss-Bonnet black brane. Treating the Gauss-Bonnet parameter $\lagb$ nonperturbatively and the momentum $q$ perturbatively, we work out $C^{(1)}$, $C^{(2)}$, confirm previous results in the literature and pave the way to a general formula for $C^{(n)}$.
\end{abstract}

%\pacs{}

\maketitle

%\tighten

%%%%%%%%%%%%%%%%%%%%%%%%%%%%%%%%%%%%%%%%%%

%\tableofcontents

\section{Introduction}\label{sect-intro}
In the research field of the AdS/CFT correspondence \cite{Maldacena:1997re,Aharony:1999ti}, one landmark result is the Kovtun-Starinets-Son bound \cite{Policastro:2001yc,Kovtun:2003wp,Buchel:2003tz,Kovtun:2004de} on $\eta/s$, the ratio of the shear viscosity to the entropy density. In the framework of the Minkowski AdS/CFT correspondence \cite{Son:2002sd}, there are several channels \cite{Policastro:2002se,Policastro:2002tn,Kovtun:2005ev,Son:2007vk} to extract the viscosity-to-entropy ratio $\eta/s$. In Ref. \cite{Kovtun:2005ev}, the spin-0 and spin-1 modes of the electromagnetic perturbations are termed as the diffusive and transverse channels respectively, while the spin-0, spin-1 and spin-2 modes of the gravitational perturbations are referred as the sound, shear and scalar channels accordingly.

Identifying the damping constant $\D$ with $\eta/(\epsilon+P)$ in the dual theory \cite{Kovtun:2003wp,Policastro:2002se}, and making use of the thermodynamic relation $\epsilon+P=Ts$, one can derive the viscosity-to-entropy ratio from the damping constant and the temperature \cite{Kovtun:2003wp,Buchel:2003tz,Son:2007vk}. For this purpose, it is crucial to compute the damping constant, which can be read off directly in the shear channel from the dispersion relation of quasinormal modes \cite{Kovtun:2003wp,Kovtun:2005ev,Brigante:2007nu}. In the Einstein gravity, the damping constant can be also inferred indirectly from the dispersion relation of quasinormal modes in the transverse channel \cite{Kovtun:2003wp,Son:2007vk,Kovtun:2005ev,Starinets:2008fb}. In both channels, corresponding to the hydrodynamic behavior of the field theory \cite{Son:2007vk}, only low-frequency quasinormal modes are relevant. In the literature, or explicitly in Refs. \cite{Kovtun:2005ev,Starinets:2008fb,Brigante:2007nu}, the customary recipe starts by rescaling $\omega\rightarrow\lambda\omega$, $q\rightarrow\lambda q$ and ends with the assumption $\omega\sim\O(q^2)$ and the result $\omega\propto q^2$. Here $\omega$, $q$ are respectively the frequency and the momentum of the quasinormal modes, and $\lambda\rightarrow0$.

As has been recognized in Refs. \cite{Kovtun:2005ev,Brigante:2007nu}, the above-mentioned rescaling implies the assumption $\omega\sim\O(q)$. This assumption is apparently at odds with the final assumption and the result. Moreover, as we pointed out in Ref. \cite{twang}, in order to get the expected result, an extra requirement should be imposed on the wave function of quasinormal modes. To fix these glitches, Ref. \cite{twang} refined the recipe by rescaling $\omega\rightarrow\lambda\omega$, $q^2\rightarrow\lambda q^2$ and assuming $\omega\sim\O(q^2)$ from beginning to end.

In Ref. \cite{twang}, the refined recipe was illustrated in two concrete examples. One example is the transverse quasinormal modes of the Schwarzschild black brane in the Einstein gravity. The other example is the shear quasinormal modes of a black brane in the Gauss-Bonnet gravity \cite{Brigante:2007nu,Brigante:2008gz}, which was studied perturbatively therein with a small Gauss-Bonnet parameter $\lagb$.

The present paper is in companion with Ref. \cite{twang}. We will proceed to study the latter example, treating the Gauss-Bonnet parameter $\lagb$ nonperturbatively. It is necessary to take $\lagb\leq1/4$, otherwise the black brane solution will not exist. For parity-even modes, the dispersion relation is dominated by $\omega\sim\O(q^2)$ in the hydrodynamic limit. The subleading-order term is of $\O(q^4)$, which we will dub as the first post-hydrodynamic order or the 1PHD order for short. To cover the 1PHD-order and higher-order terms, we will extend the rescaling rule as $\omega\rightarrow\lambda\omega_1+\lambda^2\omega_2+\cdots$, $q^2\rightarrow\lambda q^2$ and illustrate how it concretely works for the $\O(q^4)$ correction.

The outline of this paper is as follows. In the rest of this section, we will recall the line element of the Gauss-Bonnet black brane as well as its Hawking temperature. Treating the Gauss-Bonnet parameter $\lagb$ nonperturbatively in Sec. \ref{sect-GBar}, we will write down the equation of motion in the shear channel and expand it in powers of $\lambda$ with $\omega\rightarrow\lambda\omega_1+\lambda^2\omega_2+\cdots$, $q^2\rightarrow\lambda q^2$. The $\O(1)$ and $\O(\lambda)$ equations of motion will be solved in Sec. \ref{sect-hydro} to obtain the dispersion relation in the hydrodynamic limit, from which the damping constant will be read off. The 1PHD-order correction will be dug out in Sec. \ref{sect-post} from the $\O(\lambda^2)$ equation of motion. At the ends of Secs. \ref{sect-hydro} and \ref{sect-post}, our results will be compared with existing results in the literature. Further discussions will be presented in Sec. \ref{sect-sum}. Using formulas in Appendix \ref{app-form}, we will work out integrations \eqref{dg1}, \eqref{g1} at full length in Appendix \ref{app-g1} and the divergent terms of \eqref{g2v} in Appendix \ref{app-g2v}. Although the calculations are tedious, the result of Eq. \eqref{B2} is crucial for Sec. \ref{sect-hydro}, and the results of Appendices \ref{app-g1}, \ref{app-g2v} are crucial for Sec. \ref{sect-post}.

In the ($4+1$)-dimensional spacetime, a nonrotating black brane is described by the line element
\begin{equation}\label{metricr}
ds^2=\frac{r^2}{R^2}\left[-\Nsh^2f(r)dt^2+dx^2+dy^2+dz^2\right]+\frac{R^2}{r^2f(r)}dr^2.
\end{equation}
For a black brane in the Gauss-Bonnet gravity \cite{Brigante:2007nu,Brigante:2008gz,Cai:2001dz}, the function $f(r)$ and the parameter $\Nsh$ are given by
\begin{equation}\label{fGBr}
f(r)=\frac{1}{2\lagb}\left[1-\sqrt{1-4\lagb\left(1-\frac{r_0^4}{r^4}\right)}\right],~~~~\Nsh^2=\frac{1}{2}\left(1+\sqrt{1-4\lagb}\right).
\end{equation}
The Horizon is located at $r=r_0$, and the Hawking temperature is of the form
\begin{equation}
T=\frac{\Nsh r_0}{\pi R^2}.
\end{equation}
Hereafter we will take $\Nsh>0$. In the limit $\lagb\rightarrow0$, the Gauss-Bonnet gravity theory reduces to the Einstein gravity theory, and the Gauss-Bonnet black brane reduces to the Schwarzschild black brane.

Throughout this paper, to save the space and notations, we will slightly abuse the notation of the Lebesgue-Stieltjes integration. For example, in our conventions,
\begin{equation}\label{LSint}
\int_{\infty}^{\fz}f(\fz)dg(\fz)\doteq\int_{g(\infty)}^{g(\fz)}f(\fz')dg(\fz')=\int_{\infty}^{\fz}f(\fz')\frac{dg(\fz')}{d\fz'}d\fz'.
\end{equation}
When attention is paid to divergent terms, we will use $\simeq$ to denote equivalence up to convergent terms.

\begin{comment}
Following Ref. \cite{Kovtun:2005ev,Starinets:2008fb}, we introduce a dimensionless coordinate $u=r_0^2/r^2$, in terms of which Eq. \eqref{metricr} can be rewritten as
\begin{equation}
ds^2=\frac{\ptr^2}{u}\left(-f(u)dt^2+\frac{dx^2+dy^2+dz^2}{\Nsh^2}\right)+\frac{R^2}{4u^2f(u)}du^2,
\end{equation}
and equations \eqref{fSchr} and \eqref{fGBr} become
\begin{eqnarray}
&&f(u)=1-u^2,~~~~\Nsh^2=1;\\
&&f(u)=\frac{1}{2\lagb}\left[1-\sqrt{1-4\lagb\left(1-u^2\right)}\right],~~~~\Nsh^2=\frac{1}{2}\left(1+\sqrt{1-4\lagb}\right)
\end{eqnarray}
for the Schwarzschild black brane and the Gauss-Bonnet black brane respectively.
\end{comment}

\section{Shear channel of Gauss-Bonnet black brane: arbitrary $\lagb$}\label{sect-GBar}
Working with the same notations as in Ref. \cite{twang},
\begin{equation}\label{dict}
\fz^2=\frac{r^2}{r_0^2}=\frac{1}{u},~~~~\tom=\frac{R^2\omega}{r_0}=\frac{\Nsh\omega}{\pt}=2\Nsh\fw,~~~~\tq=\frac{R^2q}{r_0}=\frac{\Nsh q}{\pt}=2\Nsh\fq,
\end{equation}
we recast Eqs. \eqref{metricr}, \eqref{fGBr} to the form
\begin{eqnarray}
&&ds^2=\ptr^2\fz^2\left[-f(\fz)dt^2+\frac{dx^2+dy^2+dz^2}{\Nsh^2}\right]+\frac{R^2}{\fz^2f(\fz)}d\fz^2,\label{metricz}\\
&&f(\fz)=\frac{1}{2\lagb}\left[1-\sqrt{1-4\lagb\left(1-\frac{1}{\fz^4}\right)}\right],~~~~\Nsh^2=\frac{1}{2}\left(1+\sqrt{1-4\lagb}\right).\label{fGBz}
\end{eqnarray}
When making comparisons with Refs. \cite{Brigante:2007nu,Grozdanov:2016vgg,Grozdanov:2016fkt}, one should notice that $f(r)$ in Refs. \cite{Brigante:2007nu,Grozdanov:2016vgg,Grozdanov:2016fkt} is replaced by $r^2f(r)/R^2$ in Ref. \cite{twang} or here. What is more, $x_1,x_2,x_3$, $L$, $z$ are replaced by $y,z,x$, $R$, $\fz$ correspondingly, and $\gagb=\sqrt{1-4\lagb}$.

\begin{comment}
\begin{eqnarray}
\nonumber&&\tom=2\Nsh\fw,~~~~\tq=2\Nsh\fq,\\
\nonumber&&u=\fz^{-2},~~~~u^2-1=f(\lagb f-1),~~~~U=1-2\lagb f,\\
\nonumber&&\gagb^2=1-4\lagb,~~~~1+\gagb=\frac{2}{f(u=0)}=2\Nsh^2,\\
\nonumber&&Z_2=\frac{q}{r^2}h_{tx}+\frac{\omega}{r^2}h_{zx}=\frac{Z}{R^2}.
\end{eqnarray}
\begin{eqnarray}
\nonumber A_2&=&-\frac{2(1-4\lagb)^22\Nsh^2\fq^2}{u(-2\lagb f)(1-2\lagb f)^3}\frac{2\lagb(u^2-1)(-1-2\lagb f)-2\lagb f}{(1-4\lagb)2\Nsh^2(-2\lagb f)\fq^2+4\lagb(1-2\lagb f)^2\fw^2},\\
\nonumber B_2&=&\frac{(1-4\lagb)2\Nsh^2(2-2\lagb f)}{4u(u^2-1)(1-2\lagb f)^2}\fq^2+\frac{(2-2\lagb f)^2}{4u(u^2-1)^2}\fw^2.
\end{eqnarray}
\end{comment}

In the Gauss-Bonnet gravity, the equations of motion for fluctuations in the shear channel can be reduced to a single equation for $Z=qg^{yy}h_{ty}+\omega g^{yy}h_{xy}=R^2Z_2$. With $f$ and $\Nsh$ given by Eqs. \eqref{fGBz}, the equation is
\begin{eqnarray}\label{eomGB}
\nonumber&&\partial_{\fz}^2Z+\left[-\frac{\tq^2(1-4\lagb)}{\fz^4f(1-2\lagb f)^2}+\frac{\Nsh^{-2}\tom^2}{\fz^4f^2}\right]Z\\
&&+\left\{\frac{5}{\fz}-\frac{8\lagb}{\fz^5(1-2\lagb f)^2}+\frac{4\Nsh^{-2}\tom^2(1+2\lagb f)}{\fz^5f\left[-\tq^2(1-4\lagb)f+\Nsh^{-2}\tom^2(1-2\lagb f)^2\right]}\right\}\partial_{\fz}Z=0.
\end{eqnarray}
We derived this equation straightforwardly from the action
\begin{equation}\label{actGB}
\S=\frac{1}{16\pi G_N}\int d^5x\sqrt{-g}\left[\R-2\Lambda+\frac{1}{2}\lagb R^2\left(\R^2-4\R_{\mu\nu}\R^{\mu\nu}+\R_{\mu\nu\rho\sigma}\R^{\mu\nu\rho\sigma}\right)\right]
\end{equation}
with the cosmological constant $\Lambda=-6/R^2$. It is in agreement with the equation of motion for $Z_2$ in Refs. \cite{Grozdanov:2016vgg,Grozdanov:2016fkt}.

In the special case $\lagb=1/4$, the equation of motion simplifies drastically and can be solved by the hypergeometric function, see Ref. \cite{Grozdanov:2016fkt} for an exhaustive study. From now on we will focus on the case $\lagb<1/4$.

Near the horizon, $\fz\rightarrow1$, $f(\fz)\rightarrow0$, one can check that $\partial_{\fz}f(\fz)\rightarrow4$ and that Eq. \eqref{eomGB} is dominated by
\begin{equation}
\partial_{\fz}^2Z+\frac{4}{f}\partial_{\fz}Z+\frac{\Nsh^{-2}\tom^2}{f^2}Z=0.
\end{equation}
This equation has two classes of solutions, one for outcoming waves and one for infalling waves. At the horizon of a black brane, only infalling modes survive. In accordance with this condition, the solution to Eq. \eqref{eomGB} should take the form
\begin{equation}\label{decZ}
Z(\fz)=f^{-i\tom/(4\Nsh)}g(\fz)
\end{equation}
with $g(\fz)$ being regular at $\fz=1$. The normalization of $Z(\fz)$ can be fixed by specifying a nonzero value of $g(1)$, but we will not fix it in this paper. At spatial infinity, $\fz\rightarrow\infty$, we impose the Dirichlet boundary condition $Z(\infty)=0$, which implies $g(\infty)=0$.

Formally, in the low-frequency limit $\omega\rightarrow0$, the dispersion relation of parity-even quasinormal modes can be expanded as
\begin{equation}\label{dispexp}
\omega=C^{(1)}q^2+C^{(2)}q^4+C^{(3)}q^6+\cdots.
\end{equation}
To trace the order of $\tq$, it is convenient to introduce a book-keeping parameter $\lambda$ as in Ref. \cite{Starinets:2008fb} and rescale $\tq^2\rightarrow\lambda\tq^2$. The hydrodynamic limit corresponds to the leading-order term in Eq. \eqref{dispexp}, which has been studied by rescaling the frequency as $\tom\rightarrow\lambda\tom$ in Ref. \cite{twang} for other examples. In order to take the higher-order terms in Eq. \eqref{dispexp} into account, we will rescale the frequency and the momentum as
\begin{equation}
\tom\rightarrow\lambda\tom_1+\lambda^2\tom_2+\lambda^3\tom_3+\cdots,~~~~\tq^2\rightarrow\lambda\tq^2.
\end{equation}
Accordingly, the solution \eqref{decZ} should be rewritten as
\begin{eqnarray}\label{Zexp}
\nonumber Z(\fz,\lambda\tom_1+\lambda^2\tom_2,\lambda\tq^2)&=&f^{-i(\lambda\tom_1+\lambda^2\tom_2)/(4\Nsh)+\O(\lambda^3)}\left[g_0(\fz)+\lambda g_1(\fz)+\lambda^2g_2(\fz)+\O(\lambda^3)\right]\\
\nonumber&=&g_0+\lambda\left[g_1-\left(\frac{i\tom_1}{4\Nsh}\ln f\right)g_0\right]\\
&&+\lambda^2\left[g_2-\left(\frac{i\tom_1}{4\Nsh}\ln f\right)g_1-\left(\frac{i\tom_2}{4\Nsh}\ln f\right)g_0-\frac{1}{2}\left(\frac{\tom_1}{4\Nsh}\ln f\right)^2g_0\right]+\O(\lambda^3),
\end{eqnarray}
where $g_0(\fz)$, $g_1(\fz)$, $g_2(\fz)$ are regular at the horizon $\fz=1$. At spatial infinity, $\fz\rightarrow\infty$, $f(\fz)\rightarrow\left(1-\sqrt{1-4\lagb}\right)/(2\lagb)$, the Dirichlet boundary condition $Z(\infty)=0$ dictates
\begin{equation}\label{Dirichlet}
g_0(\infty)=\left.\left[g_1-\left(\frac{i\tom_1}{4\Nsh}\ln f\right)g_0\right]\right|_{\fz=\infty}=\left.\left[g_2-\left(\frac{i\tom_1}{4\Nsh}\ln f\right)g_1-\left(\frac{i\tom_2}{4\Nsh}\ln f\right)g_0-\frac{1}{2}\left(\frac{\tom_1}{4\Nsh}\ln f\right)^2g_0\right]\right|_{\fz=\infty}=0
\end{equation}
or equivalently $g_0(\infty)=g_1(\infty)=g_2(\infty)=0$.
\begin{comment}
Setting the upper limit of integration to $1$, we can get the divergence terms of the right hand side of Eq. \eqref{g2}, as given by Eq. \eqref{g2v}.
\begin{equation}
Z(\infty,\tom,\tq)\simeq C_0\sqrt{1-4\lagb}-C_0\sqrt{1-4\lagb}+\lambda\left(C_1\sqrt{1-4\lagb}-C_1\sqrt{1-4\lagb}\right)+\lambda^2\left(C_2\sqrt{1-4\lagb}-C_2\sqrt{1-4\lagb}\right).
\end{equation}
Thus the Dirichlet boundary condition $E_x(0)=0$ are satisfied by fixing the values $-C_0\sqrt{1-4\lagb}=-C_0\sqrt{1-4\lagb}$, $-C_1\sqrt{1-4\lagb}=-C_1\sqrt{1-4\lagb}$, $-C_2\sqrt{1-4\lagb}=-C_2\sqrt{1-4\lagb}$.
\end{comment}

In the case $\lagb<1/4$, expanding Eq. \eqref{eomGB} in powers of $\lambda$,
\begin{comment}
\begin{eqnarray}
\nonumber&&\partial_{\fz}^2Z+\left[5-\frac{8\lagb}{\fz^4\left(1-4\lagb+4\lagb\fz^{-4}\right)}\right]\frac{1}{\fz}\partial_{\fz}Z-\lambda\Biggl[\frac{16\Nsh^{-2}\tom_1^2\lagb^2\left(2-\sqrtz\right)}{\tq^2(1-4\lagb)\fz^5\left(1-\sqrtz\right)^2}\partial_{\fz}Z\\
\nonumber&&+\frac{2\tq^2\lagb(1-4\lagb)}{\fz^4\left(1-4\lagb+4\lagb\fz^{-4}\right)\left(1-\sqrtz\right)}Z\Biggr]+\lambda^2\Biggl[-\frac{32\Nsh^{-2}\tom_1\tom_2\lagb^2\left(2-\sqrtz\right)}{\tq^2(1-4\lagb)\fz^5\left(1-\sqrtz\right)^2}\partial_{\fz}Z\\
\nonumber&&-\frac{32\Nsh^{-4}\tom_1^4\lagb^3\left(1-4\lagb+4\lagb\fz^{-4}\right)\left(2-\sqrtz\right)}{\tq^4(1-4\lagb)^2\fz^5\left(1-\sqrtz\right)^3}\partial_{\fz}Z+\frac{4\Nsh^{-2}\tom_1^2\lagb^2}{\fz^4\left(1-\sqrtz\right)^2}Z\Biggr]\\
&&+\O(\lambda^3)=0,
\end{eqnarray}
\end{comment}
and substituting the solution \eqref{Zexp} into it, we put the equation of motion in order,
\begin{eqnarray}\label{eomGBexp}
\nonumber&&\partial_{\fz}^2g_0+\left[5-\frac{8\lagb}{\fz^4\left(1-4\lagb+4\lagb\fz^{-4}\right)}\right]\frac{1}{\fz}\partial_{\fz}g_0\\
\nonumber&&+\lambda\Biggl\{\partial_{\fz}^2\left[g_1-\left(\frac{i\tom_1}{4\Nsh}\ln f\right)g_0\right]+\left[5-\frac{8\lagb}{\fz^4\left(1-4\lagb+4\lagb\fz^{-4}\right)}\right]\frac{1}{\fz}\partial_{\fz}\left[g_1-\left(\frac{i\tom_1}{4\Nsh}\ln f\right)g_0\right]\\
\nonumber&&-\frac{16\Nsh^{-2}\tom_1^2\lagb^2\left(2-\sqrtz\right)}{\tq^2(1-4\lagb)\fz^5\left(1-\sqrtz\right)^2}\partial_{\fz}g_0-\frac{2\tq^2\lagb(1-4\lagb)}{\fz^4\left(1-4\lagb+4\lagb\fz^{-4}\right)\left(1-\sqrtz\right)}g_0\Biggr\}\\
\nonumber&&+\lambda^2\Biggl\{\partial_{\fz}^2\left[g_2-\left(\frac{i\tom_1}{4\Nsh}\ln f\right)g_1-\left(\frac{i\tom_2}{4\Nsh}\ln f\right)g_0-\frac{1}{2}\left(\frac{\tom_1}{4\Nsh}\ln f\right)^2g_0\right]\\
\nonumber&&+\left[5-\frac{8\lagb}{\fz^4\left(1-4\lagb+4\lagb\fz^{-4}\right)}\right]\frac{1}{\fz}\partial_{\fz}\left[g_2-\left(\frac{i\tom_1}{4\Nsh}\ln f\right)g_1-\left(\frac{i\tom_2}{4\Nsh}\ln f\right)g_0-\frac{1}{2}\left(\frac{\tom_1}{4\Nsh}\ln f\right)^2g_0\right]\\
\nonumber&&-\frac{16\Nsh^{-2}\tom_1^2\lagb^2\left(2-\sqrtz\right)}{\tq^2(1-4\lagb)\fz^5\left(1-\sqrtz\right)^2}\partial_{\fz}\left[g_1-\left(\frac{i\tom_1}{4\Nsh}\ln f\right)g_0\right]\\
\nonumber&&-\frac{2\tq^2\lagb(1-4\lagb)}{\fz^4\left(1-4\lagb+4\lagb\fz^{-4}\right)\left(1-\sqrtz\right)}\left[g_1-\left(\frac{i\tom_1}{4\Nsh}\ln f\right)g_0\right]\\
\nonumber&&-\frac{32\Nsh^{-2}\tom_1\tom_2\lagb^2\left(2-\sqrtz\right)}{\tq^2(1-4\lagb)\fz^5\left(1-\sqrtz\right)^2}\partial_{\fz}g_0\\
%\end{eqnarray}
%\begin{eqnarray}
\nonumber&&-\frac{32\Nsh^{-4}\tom_1^4\lagb^3\left(1-4\lagb+4\lagb\fz^{-4}\right)\left(2-\sqrtz\right)}{\tq^4(1-4\lagb)^2\fz^5\left(1-\sqrtz\right)^3}\partial_{\fz}g_0+\frac{4\Nsh^{-2}\tom_1^2\lagb^2}{\fz^4\left(1-\sqrtz\right)^2}g_0\Biggr\}\\
&&+\O(\lambda^3)=0.
\end{eqnarray}
By definition, functions $g_0(\fz)$, $g_1(\fz)$ and $g_2(\fz)$ are free of $\lambda$, and the equation of motion is expected to hold in each order of $\lambda$. Hence this equation can be solved order by order. The $\O(1)$ and $\O(\lambda)$ equations will be studied in Sec. \ref{sect-hydro}, while the $\O(\lambda^2)$ equation will be investigated in Sec. \ref{sect-post}.

\section{Hydrodynamic limit}\label{sect-hydro}
To sudy the hydrodynamic behavior of quasinormal modes in the shear channel, we collect the $\O(1)$ and $\O(\lambda)$ terms in Eq. \eqref{eomGBexp}, and obtain two equations:
\begin{eqnarray}
&&\frac{1}{\fz^5\sqrtz}\partial_{\fz}\left(\fz^5\sqrtz\partial_{\fz}g_0\right)=0,\label{ddg0}\\
\nonumber&&\frac{1}{\fz^5\sqrtz}\partial_{\fz}\left\{\fz^5\sqrtz\partial_{\fz}\left[g_1-\left(\frac{i\tom_1}{4\Nsh}\ln f\right)g_0\right]\right\}\\
\nonumber&&-\frac{16\Nsh^{-2}\tom_1^2\lagb^2\left(2-\sqrtz\right)}{\tq^2(1-4\lagb)\fz^5\left(1-\sqrtz\right)^2}\partial_{\fz}g_0-\frac{2\tq^2\lagb(1-4\lagb)}{\fz^4\left(1-4\lagb+4\lagb\fz^{-4}\right)\left(1-\sqrtz\right)}g_0=0.\\
\label{ddg1}
\end{eqnarray}
Making use of Eqs. \eqref{dz} and the Dirichlet boundary condition \eqref{Dirichlet}, we find Eq. \eqref{ddg0} can be solved as
\begin{eqnarray}\label{g0}
\nonumber\partial_{\fz}g_0&=&\frac{C_0}{\fz^5\sqrtz},\\
g_0&=&-\frac{C_0}{8\lagb}\left(\sqrtz-\sqrt{1-4\lagb}\right).
\end{eqnarray}
In the same way, we obtain the first integral of Eq. \eqref{ddg1},
\begin{eqnarray}\label{dg1}
\nonumber&&\partial_{\fz}\left[g_1-\left(\frac{i\tom_1}{4\Nsh}\ln f\right)g_0\right]\\
\nonumber&=&\frac{1}{\fz^5\sqrtz}\Biggl[\int_{\infty}^{\fz}\frac{16C_0\Nsh^{-2}\tom_1^2\lagb^2\left(2-\sqrtz\right)}{\tq^2(1-4\lagb)\fz^5\left(1-\sqrtz\right)^2}d\fz\\
\nonumber&&-\int_{\infty}^{\fz}\frac{C_0\tq^2(1-4\lagb)\fz\left(\sqrtz-\sqrt{1-4\lagb}\right)}{4\sqrtz\left(1-\sqrtz\right)}d\fz+C_1\Biggr]\\
&=&\frac{1}{\fz^5\sqrtz}\times\eqref{B1},
\end{eqnarray}
and its second integral
\begin{eqnarray}\label{g1}
\nonumber&&g_1-\left(\frac{i\tom_1}{4\Nsh}\ln f\right)g_0\\
\nonumber&=&\int_{\infty}^{\fz}\partial_{\fz}\left[g_1-\left(\frac{i\tom_1}{4\Nsh}\ln f\right)g_0\right]d\fz\\
&=&\eqref{B2}
\end{eqnarray}
in accordance with the Dirichlet boundary condition \eqref{Dirichlet}. It is tricky to work out the above integrations. After presenting some useful formulas in Appendix \ref{app-form}, we accomplish the task in Appendix \ref{app-g1}. See the results of Eqs. \eqref{B1}, \eqref{B2}.

As mentioned above, the regularity of $g(\fz)$ at $\fz=1$ demands the finiteness of $g_0(1)$ and $g_1(1)$. Apparently $g_0(1)$ is finite,
\begin{equation}\label{g0v}
g_0(1)=-\frac{C_0}{8\lagb}\left(1-\sqrt{1-4\lagb}\right).
\end{equation}
The finiteness of $g_1(1)$ can be studied by taking the limit $\fz\rightarrow1$ of Eq. \eqref{B2}. In this limit, the last term at the end of Eq. \eqref{B2} goes to zero according to Eq. \eqref{limln2}, and the divergent part of Eq. \eqref{B2} turns out to be
\begin{eqnarray}
\nonumber\left.\eqref{B2}\right|_{\fz=1}&\simeq&-\left.\frac{C_0\Nsh^{-2}\tom_1^2}{4\tq^2(1-4\lagb)}\ln\frac{1-\sqrtz}{1-\sqrt{1-4\lagb}}\right|_{\fz=1}\\
&\simeq&-\frac{C_0\Nsh^{-2}\tom_1^2}{4\tq^2(1-4\lagb)}\left.(\ln f)\right|_{\fz=1}.
\end{eqnarray}
Therefore, the potentially divergent terms in $g_1(1)$ are
%\begin{eqnarray}
%\nonumber g_1(1)&=&\left(\frac{i\tom_1}{4\Nsh}\ln f\right)g_0(1)+\left.\eqref{B2}\right|_{\fz=1}\\
%\nonumber&\simeq&-\frac{iC_0\tom_1}{32\Nsh\lagb}\left(1-\sqrt{1-4\lagb}\right)\left.(\ln f)\right|_{\fz=1}-\frac{C_0\Nsh^{-2}\tom_1^2}{4\tq^2(1-4\lagb)}\left.(\ln f)\right|_{\fz=1}.
%\end{eqnarray}
\begin{equation}
g_1(1)\simeq-\frac{iC_0\tom_1}{32\Nsh\lagb}\left(1-\sqrt{1-4\lagb}\right)\left.(\ln f)\right|_{\fz=1}-\frac{C_0\Nsh^{-2}\tom_1^2}{4\tq^2(1-4\lagb)}\left.(\ln f)\right|_{\fz=1}.
\end{equation}
Here we care only about terms that are divergent at $\fz=1$, so we have use $\simeq$ to denote equivalence up to finite terms. The value of $\Nsh^2$ is given by Eqs. \eqref{fGBz}. Note that $\ln f$ is divergent at $\fz=1$, the finiteness of $g_1(1)$ requires the cancellation of its coefficients, that is
%\begin{equation}
%-\frac{iC_0\tom_1}{32\Nsh\lagb}\left(1-\sqrt{1-4\lagb}\right)-\frac{C_0\Nsh^{-2}\tom_1^2}{4\tq^2(1-4\lagb)}=0.
%\end{equation}
%That is,
\begin{equation}\label{tomq}
\frac{i}{4\Nsh}+\frac{\tom_1}{\tq^2(1-4\lagb)}=0.
\end{equation}
%\begin{equation}
%-\frac{iC_0\omega_1}{8\pt}\frac{1}{1+\sqrt{1-4\lagb}}-\frac{C_0\Nsh^{-2}\omega_1^2}{4q^2(1-4\lagb)}=0.
%\end{equation}
Trading $\tom_1$ to $\omega_1$, and $\tq$ to $q$ with the help of Eqs. \eqref{dict}, we find the cancellation condition above becomes
\begin{equation}
\omega_1=-iq^2\frac{1-4\lagb}{4\pt}.
\end{equation}
Comparing it with the dispersion relation $\omega=-i\mathcal{D}q^2$ of damped hydrodynamic shear flow \cite{Kovtun:2003wp,Kovtun:2005ev,Brigante:2007nu}, we get the damping constant
\begin{equation}
\mathcal{D}=\frac{1-4\lagb}{4\pt}.
\end{equation}
Identifying the damping constant $\D$ with $\eta/(\epsilon+P)$ in the dual theory \cite{Kovtun:2003wp,Policastro:2002se}, and making use of the thermodynamic relation $\epsilon+P=Ts$, one can derive the viscosity-to-entropy ratio from the damping constant and the temperature by $\eta/s=T\mathcal{D}$ \cite{Kovtun:2003wp,Buchel:2003tz,Son:2007vk}. As a result, the viscosity-to-entropy ratio of the Gauss-Bonnet holographic liquid is $\eta/s=(1-4\lagb)/(4\pi)$. This result is nonperturbative in $\lagb$. It has been obtained in Ref. \cite{Brigante:2007nu} from the Kubo formula, see Eq. (3.23) therein. Here we confirmed the result directly in the shear channel.

\section{1PHD-order correction}\label{sect-post}
In the full dispersion relation Eq. \eqref{dispexp}, the $\O(q^2)$ term dominates in the hydrodynamic limit. After this term comes the $\O(q^4)$ term. We use the first-post-hydrodynamic-order (1PHD-order) correction to refer to such a subleading-order term. To calculate the 1PHD-order correction, we should take care of the $\O(\lambda^2)$ terms in Eq. \eqref{eomGBexp}, which give a lengthy equation,
\begin{eqnarray}
\nonumber&&\frac{1}{\fz^5\sqrtz}\partial_{\fz}\Biggl\{\fz^5\sqrtz\partial_{\fz}\Biggl[g_2-\left(\frac{i\tom_1}{4\Nsh}\ln f\right)g_1-\left(\frac{i\tom_2}{4\Nsh}\ln f\right)g_0\\
\nonumber&&-\frac{1}{2}\left(\frac{\tom_1}{4\Nsh}\ln f\right)^2g_0\Biggr]\Biggr\}-\frac{16\Nsh^{-2}\tom_1^2\lagb^2\left(2-\sqrtz\right)}{\tq^2(1-4\lagb)\fz^5\left(1-\sqrtz\right)^2}\partial_{\fz}\left[g_1-\left(\frac{i\tom_1}{4\Nsh}\ln f\right)g_0\right]\\
\nonumber&&-\frac{2\tq^2\lagb(1-4\lagb)}{\fz^4\left(1-4\lagb+4\lagb\fz^{-4}\right)\left(1-\sqrtz\right)}\left[g_1-\left(\frac{i\tom_1}{4\Nsh}\ln f\right)g_0\right]\\
\nonumber&&-\frac{32\Nsh^{-2}\tom_1\tom_2\lagb^2\left(2-\sqrtz\right)}{\tq^2(1-4\lagb)\fz^5\left(1-\sqrtz\right)^2}\partial_{\fz}g_0\\
\nonumber&&-\frac{32\Nsh^{-4}\tom_1^4\lagb^3\left(1-4\lagb+4\lagb\fz^{-4}\right)\left(2-\sqrtz\right)}{\tq^4(1-4\lagb)^2\fz^5\left(1-\sqrtz\right)^3}\partial_{\fz}g_0+\frac{4\Nsh^{-2}\tom_1^2\lagb^2}{\fz^4\left(1-\sqrtz\right)^2}g_0=0.\\
\end{eqnarray}
Similar to the previous section, in accordance with the Dirichlet boundary condition \eqref{Dirichlet}, formally we can rearrange the first integral of this equation as
\begin{eqnarray}\label{dg2}
\nonumber&&\partial_{\fz}\left[g_2-\left(\frac{i\tom_1}{4\Nsh}\ln f\right)g_1-\left(\frac{i\tom_2}{4\Nsh}\ln f\right)g_0-\frac{1}{2}\left(\frac{\tom_1}{4\Nsh}\ln f\right)^2g_0\right]\\
\nonumber&=&\frac{1}{\fz^5\sqrtz}\Biggl[\int_{\infty}^{\fz}\frac{16\Nsh^{-2}\tom_1^2\lagb^2\left(2-\sqrtz\right)}{\tq^2(1-4\lagb)\fz^5\left(1-\sqrtz\right)^2}\times\eqref{B1}d\fz\\
\nonumber&&+\int_{\infty}^{\fz}\frac{2\tq^2\lagb(1-4\lagb)\fz}{\sqrtz\left(1-\sqrtz\right)}\times\eqref{B2}d\fz\\
\nonumber&&+\int_{\infty}^{\fz}\frac{32C_0\Nsh^{-2}\tom_1\tom_2\lagb^2\left(2-\sqrtz\right)}{\tq^2(1-4\lagb)\fz^5\left(1-\sqrtz\right)^2}d\fz\\
\nonumber&&+\int_{\infty}^{\fz}\frac{32C_0\Nsh^{-4}\tom_1^4\lagb^3\left(1-4\lagb+4\lagb\fz^{-4}\right)\left(2-\sqrtz\right)}{\tq^4(1-4\lagb)^2\fz^5\left(1-\sqrtz\right)^3}d\fz\\
\nonumber&&+\int_{\infty}^{\fz}\frac{C_0\Nsh^{-2}\tom_1^2\lagb\fz\sqrtz}{2\left(1-\sqrtz\right)^2}\left(\sqrtz-\sqrt{1-4\lagb}\right)d\fz+C_2\Biggr]\\
&=&\frac{1}{\fz^5\sqrtz}\times\eqref{C1}
\end{eqnarray}
and its second integral as
\begin{eqnarray}\label{g2}
\nonumber&&g_2-\left(\frac{i\tom_1}{4\Nsh}\ln f\right)g_1-\left(\frac{i\tom_2}{4\Nsh}\ln f\right)g_0-\frac{1}{2}\left(\frac{\tom_1}{4\Nsh}\ln f\right)^2g_0\\
\nonumber&=&\int_{\infty}^{\fz}\partial_{\fz}\left[g_2-\left(\frac{i\tom_1}{4\Nsh}\ln f\right)g_1-\left(\frac{i\tom_2}{4\Nsh}\ln f\right)g_0-\frac{1}{2}\left(\frac{\tom_1}{4\Nsh}\ln f\right)^2g_0\right]d\fz\\
\nonumber&=&\frac{1}{8\lagb}\int_{\infty}^{\fz}\eqref{C1}d\left(1-\sqrtz\right)\\
%\nonumber&=&\frac{1}{8\lagb}\left(1-\sqrtz\right)\times\eqref{C1}-\frac{1}{8\lagb}\left(1-\sqrt{1-4\lagb}\right)\times\left.\eqref{C1}\right|_{\fz=\infty}\\
%\nonumber&&-\frac{1}{8\lagb}\Biggl[\int_{\infty}^{\fz}\frac{16\Nsh^{-2}\tom_1^2\lagb^2\left(2-\sqrtz\right)}{\tq^2(1-4\lagb)\fz^5\left(1-\sqrtz\right)}\times\eqref{B1}d\fz\\
%\nonumber&&+\int_{\infty}^{\fz}\frac{2\tq^2\lagb(1-4\lagb)\fz}{\sqrtz}\times\eqref{B2}d\fz+\int_{\infty}^{\fz}\frac{32C_0\Nsh^{-2}\tom_1\tom_2\lagb^2\left(2-\sqrtz\right)}{\tq^2(1-4\lagb)\fz^5\left(1-\sqrtz\right)}d\fz\\
%\nonumber&&+\int_{\infty}^{\fz}\frac{32C_0\Nsh^{-4}\tom_1^4\lagb^3\left(1-4\lagb+4\lagb\fz^{-4}\right)\left(2-\sqrtz\right)}{\tq^4(1-4\lagb)^2\fz^5\left(1-\sqrtz\right)^2}d\fz\\
%\nonumber&&+\int_{\infty}^{\fz}\frac{C_0\Nsh^{-2}\tom_1^2\lagb\fz\sqrtz}{2\left(1-\sqrtz\right)}\left(\sqrtz-\sqrt{1-4\lagb}\right)d\fz\Biggr]\\
&=&\frac{1}{8\lagb}\left(1-\sqrtz\right)\times\eqref{C1}-\frac{1}{8\lagb}\times\eqref{C2}.
\end{eqnarray}
In the third line of Eq. \eqref{g2}, we have applied Eqs. \eqref{dz}. In the fourth line, we have made integration by parts and used $\left.\eqref{C1}\right|_{\fz=\infty}=0$.

Evaluated at the horizon $\fz=1$ and multiplied by $-8\lagb$, Eq. \eqref{g2} takes the form
\begin{eqnarray}\label{g2v}
\nonumber-8\lagb\left.\left[g_2-\left(\frac{i\tom_1}{4\Nsh}\ln f\right)g_1-\left(\frac{i\tom_2}{4\Nsh}\ln f\right)g_0-\frac{1}{2}\left(\frac{\tom_1}{4\Nsh}\ln f\right)^2g_0\right]\right|_{\fz=1}&=&-\left.\left(1-\sqrtz\right)\times\eqref{C1}\right|_{\fz=1}\\
&&+\left.\eqref{C2}\right|_{\fz=1}.
\end{eqnarray}
It is tedious to derive the divergent terms on both sides of Eq. \eqref{g2v}. The heavy details are relegated to Appendix \ref{app-g2v}, and the final results are given by the last steps of Eqs. \eqref{g2vleft}, \eqref{g2vrighta}, \eqref{g2vrightb}. Equating \eqref{g2vleft} to the sum of \eqref{g2vrighta} and \eqref{g2vrightb}, we write down
\begin{eqnarray}\label{g2div}
\nonumber&&\Biggl\{-\frac{iC_0\Nsh^{-3}\tom_1^3\lagb}{2\tq^2(1-4\lagb)}\left[-\ln\frac{1-\sqrt{1-4\lagb}}{2\lagb}+1+\frac{1}{2}\left(1-\sqrt{1-4\lagb}\right)^2\right]-\frac{iC_1\Nsh^{-1}\tom_1}{4}\left(1-\sqrt{1-4\lagb}\right)\\
\nonumber&&-\frac{iC_0\Nsh^{-1}\tom_1\tq^2\lagb\sqrt{1-4\lagb}}{8\left(1+\sqrt{1-4\lagb}\right)}-\frac{iC_0\Nsh^{-1}\tom_2}{4}\left(1-\sqrt{1-4\lagb}\right)\Biggr\}\left.(\ln f)\right|_{\fz=1}\\
\nonumber&&+\left[\frac{C_0\Nsh^{-2}\tom_1^2}{32}\left(1-\sqrt{1-4\lagb}\right)-\frac{iC_0\Nsh^{-3}\tom_1^3\lagb}{2\tq^2(1-4\lagb)}\right]\left.(\ln f)^2\right|_{\fz=1}\\
\nonumber&=&-\frac{C_0\Nsh^{-2}\tom_1^2\lagb}{4\left(1+\sqrt{1-4\lagb}\right)}\left.(\ln f)\right|_{\fz=1}+\Biggl[\frac{4C_0\Nsh^{-4}\tom_1^4\lagb^2}{\tq^4(1-4\lagb)\left(1-\sqrt{1-4\lagb}\right)}+\frac{2C_1\Nsh^{-2}\tom_1^2\lagb}{\tq^2(1-4\lagb)}\\
\nonumber&&+\frac{C_0\Nsh^{-2}\tom_1^2\lagb}{8\left(1+\sqrt{1-4\lagb}\right)}\ln\frac{16}{1-4\lagb}+\frac{C_0\Nsh^{-2}\tom_1^2\lagb}{4\sqrt{1-4\lagb}\left(1+\sqrt{1-4\lagb}\right)}+\frac{4C_0\Nsh^{-2}\tom_1\tom_2\lagb}{\tq^2(1-4\lagb)}\Biggr]\left.(\ln f)\right|_{\fz=1}\\
&&-\frac{C_0\Nsh^{-2}\tom_1^2\lagb}{8\left(1+\sqrt{1-4\lagb}\right)}\left.(\ln f)^2\right|_{\fz=1}.
\end{eqnarray}
Terms proportional to $C_1\left.(\ln f)\right|_{\fz=1}$, $C_0\left.(\ln f)^2\right|_{\fz=1}$, $C_0\left.(\ln f)\right|_{\fz=1}$ are independent. As a result, the finiteness of $g_2(1)$ demands that their coefficients should cancel independently in this equation. The cancellation of $C_1\left.(\ln f)\right|_{\fz=1}$ terms leads to Eq. \eqref{tomq}, and the same with $C_0\left.(\ln f)^2\right|_{\fz=1}$ terms. In contrast, the cancellation of $C_0\left.(\ln f)\right|_{\fz=1}$ terms leads to a constraint on $\tom_2$. After substitution of Eq. \eqref{tomq} and some straightforward algebras, we finally arrive at the result
\begin{eqnarray}
\nonumber\tom_2&=&\frac{i\tom_1^2}{8\Nsh}\left(4\Nsh^4+\ln\frac{1-4\lagb}{16\Nsh^4}\right)\\
&=&-\frac{i\tq^4(1-4\lagb)^2}{128\Nsh^3}\left(4\Nsh^4+\ln\frac{1-4\lagb}{16\Nsh^4}\right),
\end{eqnarray}
or getting rid of tildes with the help of Eqs. \eqref{dict},
\begin{equation}
\omega_1+\omega_2=-iq^2\frac{1-4\lagb}{4\pt}-\frac{iq^4(1-4\lagb)^2}{128(\pt)^3}\left(4\Nsh^4+\ln\frac{1-4\lagb}{16\Nsh^4}\right).
\end{equation}
Translated into $\fw$, $\fq$ and $\gagb$ via Eqs. \eqref{dict}, \eqref{fGBz} and $\gagb=\sqrt{1-4\lagb}$, this result coincides with Eq. (4.17) in Ref. \cite{Grozdanov:2016vgg} or equivalently Eq. (2.49) in Ref. \cite{Grozdanov:2016fkt},
\begin{equation}\label{dispGS}
\fw=-i\frac{\gagb^2}{2}\fq^2-i\frac{\gagb^4}{16}\left[(1+\gagb)^2+2\ln\frac{\gagb}{2(1+\gagb)}\right]\fq^4+\cdots.
\end{equation}

As mentioned honestly around Eq. (2.52) in Ref. \cite{Grozdanov:2016fkt}, in order to obtain this hydrodynamic dispersion relation, they needed to find $Z_2$ with both $\fw\sim\mu\ll1$ and $\fq\sim\mu\ll1$ scaling the same way, but use the scaling $\fw\sim\mu^2\ll1$ and $\fq\sim\mu\ll1$ to extract the diffusive pole. Here we verified the same result but avoided assuming $\fw\sim\fq$.

It is remarkable that in the limit $\lagb\rightarrow0$, which corresponds to the Schwarzschild black brane in the Einstein gravity, the dispersion relation Eq. \eqref{dispGS} reduces to
\begin{equation}\label{dispBRSSS}
\fw=-\frac{i}{2}\fq^2-\frac{i}{4}\left(1-\ln2\right)\fq^4+\cdots
\end{equation}
in agreement with Ref. \cite{Baier:2007ix}. In Ref. \cite{Baier:2007ix}, the dispersion relation Eq. \eqref{dispBRSSS} was obtained by solving the differential equation for $G(u)=(1-u)^{i\fw/2}(uh_{ty})'/(\pt R)^2$ assuming $\fw\sim\fq^2$ consistently.

\begin{comment}
\begin{figure}
\centering
\includegraphics[width=0.3\textwidth]{tom2.eps}\\
\caption{(color online). The volume \eqref{VdS} outside dS horizon in terms of the cutoff radius $r_c$. The black solid line, the purple dotted line and the blue dashed line depict numerical result of the first, the third and the last lines of equation \eqref{VdS} respectively.}\label{fig-dS}
\end{figure}
\end{comment}

\section{Summary}\label{sect-sum}
In the fluid/gravity duality, the dispersion relation of quasinormal modes provides a rich source for extracting transport coefficients of the dual fluid. For the Gauss-Bonnet black brane with an arbitrary Gauss-Bonnet parameter $\lagb$, the shear quasinormal modes obey the equation of motion \eqref{eomGB}. In this paper, we have applied the refined recipe \cite{twang} to compute the dispersion relation to the first post-hydrodynamic order $\omega=C^{(1)}q^2+C^{(2)}q^4+\cdots$, nonperturbatively in  $\lagb$. The dispersion relation in the hydrodynamic limit $\omega_1=C^{(1)}q^2$ has been derived in Sec. \ref{sect-hydro}, agreeing well with the celebrated result obtained in Ref. \cite{Brigante:2007nu} from the Kubo formula. The 1PHD-order term $\omega_2=C^{(2)}q^4$ has been computed in Sec. \ref{sect-post}, in coincidence with the result of Ref. \cite{Grozdanov:2016vgg,Grozdanov:2016fkt} which have assumed firstly $\fw\sim\O(\fq)$ and finally $\fw\sim\O(\fq^2)$.

Following the same recipe, it is straightforward to extend our calculation to 2PDH and higher order. Ultimately, it will lead to a general formula for $C^{(n)}$ in the dispersion relation $\omega=\sum_n C^{(n)}q^{2n}$. In general, to get $C^{(n)}$ one should work out $g_0(\fz)$, $g_1(\fz)$, ..., $g_{n-1}(\fz)$. Owing to the complication of Eq. \eqref{eomGB}, this will be a tremendous but surmountable challenge for future study.

\begin{acknowledgments}
This work is supported by the National Natural Science Foundation of China (Grant No. 91536218). %The author thanks Li Li for collaboration in the early stage of this work.
\end{acknowledgments}

\appendix

\section{Useful formulas}\label{app-form}
In our calculations throughout this paper, we find it is helpful to slightly abuse the notation of the Lebesgue-Stieltjes integration as Eq. \eqref{LSint} and make use of the following formulas:
\begin{eqnarray}\label{dz}
\nonumber&&\ln\left(1-\sqrtz\right)=\ln f+\ln(2\lagb),\\
\nonumber&&\frac{(1-4\lagb)\fz}{\sqrtz}d\fz=\frac{1}{2}d\sqrt{\fz^4-4\lagb\fz^4+4\lagb},\\
\nonumber&&\frac{1}{\fz^5}d\fz=-\frac{1}{8\lagb}\sqrtz d\sqrtz,\\
\nonumber&&\int_{\infty}^1\left[\int_{\infty}^{\fz}\frac{1}{1-\sqrtz}\left(\cdots\right)d\fz\right]d\sqrtz=\int_{\infty}^1\left(\cdots\right)d\fz,\\
&&\int_{\infty}^1\left[\frac{1}{\fz^5\sqrtz}\int_{\infty}^{\fz}\frac{1}{1-\sqrtz}\left(\cdots\right)d\fz\right]d\fz=-\frac{1}{8\lagb}\int_{\infty}^1\left(\cdots\right)d\fz.
\end{eqnarray}
Here $\cdots$ denotes a function that tends to zero as $\fz\rightarrow\infty$, and $f$ is given by Eqs. \eqref{fGBz}.

In the case with $0<\lagb<1/4$, we introduce the notation $a=\sqrt{4\lagb/(1-4\lagb)}$ and the variable $\tu=a\fz^{-2}$, then the following integral formulas are useful in our calculations:
\begin{eqnarray}\label{duln}
\nonumber\int\frac{1}{\sqrt{1+\tu^2}}d\tu&=&\ln\left(\tu+\sqrt{1+\tu^2}\right),\\
\nonumber\int\frac{1}{\sqrt{1+\tu^2}\pm1}d\tu&=&\ln\left(\tu+\sqrt{1+\tu^2}\right)-\frac{\tu}{\sqrt{1+\tu^2}\pm1},\\
%\nonumber\int\frac{1}{\sqrt{1+a^2}-\sqrt{1+\tu^2}}d\tu&=&-\ln\left(\tu+\sqrt{1+\tu^2}\right)+\frac{\sqrt{1+a^2}}{2a}\left(\ln\frac{a+\tu}{a-\tu}+\ln\frac{a\sqrt{1+\tu^2}+\tu\sqrt{1+a^2}}{a\sqrt{1+\tu^2}-\tu\sqrt{1+a^2}}\right),\\
\nonumber\int\frac{1}{\sqrt{1+a^2}-\sqrt{1+\tu^2}}d\tu&=&-\ln\left(\tu+\sqrt{1+\tu^2}\right)+\frac{\sqrt{1+a^2}}{2a}\ln\frac{(a+\tu)\left(a\sqrt{1+\tu^2}+\tu\sqrt{1+a^2}\right)}{(a-\tu)\left(a\sqrt{1+\tu^2}-\tu\sqrt{1+a^2}\right)},\\
%\nonumber\int\frac{1}{\left(\sqrt{1+a^2}-\sqrt{1+\tu^2}\right)^2}d\tu&=&\frac{x\sqrt{1+a^2}}{a^2\left(\sqrt{1+a^2}-\sqrt{1+\tu^2}\right)}+\frac{1}{2a^3}\left(\ln\frac{a+\tu}{a-\tu}+\ln\frac{a\sqrt{1+\tu^2}+\tu\sqrt{1+a^2}}{a\sqrt{1+\tu^2}-\tu\sqrt{1+a^2}}\right),\\
\int\frac{1}{\left(\sqrt{1+a^2}-\sqrt{1+\tu^2}\right)^2}d\tu&=&\frac{\tu\sqrt{1+a^2}}{a^2\left(\sqrt{1+a^2}-\sqrt{1+\tu^2}\right)}+\frac{1}{2a^3}\ln\frac{(a+\tu)\left(a\sqrt{1+\tu^2}+\tu\sqrt{1+a^2}\right)}{(a-\tu)\left(a\sqrt{1+\tu^2}-\tu\sqrt{1+a^2}\right)}.
\end{eqnarray}
For the other case $\lagb<0$, we can introduce $a=\sqrt{-4\lagb/(1-4\lagb)}$, $\tu=a\fz^{-2}$ and make use of the integral formulas
\begin{eqnarray}\label{duarc}
\nonumber\int\frac{1}{\sqrt{1-\tu^2}}d\tu&=&\arcsin\tu,\\
\nonumber\int\frac{1}{\sqrt{1-\tu^2}\pm1}d\tu&=&\arcsin\tu-\frac{\tu}{\sqrt{1-\tu^2}\pm1},\\
%\nonumber\int\frac{1}{\sqrt{1-a^2}-\sqrt{1-\tu^2}}d\tu&=&-\arcsin\tu-\frac{\sqrt{1-a^2}}{a}\left(\arctanh\frac{\tu}{a}+\arctanh\frac{\tu\sqrt{1-a^2}}{a\sqrt{1-\tu^2}}\right),\\
\nonumber\int\frac{1}{\sqrt{1-a^2}-\sqrt{1-\tu^2}}d\tu&=&-\arcsin\tu-\frac{\sqrt{1-a^2}}{2a}\ln\frac{(a+\tu)\left(a\sqrt{1+\tu^2}+\tu\sqrt{1+a^2}\right)}{(a-\tu)\left(a\sqrt{1+\tu^2}-\tu\sqrt{1+a^2}\right)},\\
%\nonumber\int\frac{1}{\left(\sqrt{1-a^2}-\sqrt{1-\tu^2}\right)^2}d\tu&=&-\frac{x\sqrt{1-a^2}}{a^2\left(\sqrt{1-a^2}-\sqrt{1-\tu^2}\right)}+\frac{1}{a^3}\left(\arctanh\frac{\tu}{a}+\arctanh\frac{\tu\sqrt{1-a^2}}{a\sqrt{1-\tu^2}}\right),\\
\int\frac{1}{\left(\sqrt{1-a^2}-\sqrt{1-\tu^2}\right)^2}d\tu&=&-\frac{\tu\sqrt{1-a^2}}{a^2\left(\sqrt{1-a^2}-\sqrt{1-\tu^2}\right)}+\frac{1}{2a^3}\ln\frac{(a+\tu)\left(a\sqrt{1-\tu^2}+\tu\sqrt{1-a^2}\right)}{(a-\tu)\left(a\sqrt{1-\tu^2}-\tu\sqrt{1-a^2}\right)}.
\end{eqnarray}

For both cases, we can perform the following five integrations to get the same results:
\begin{eqnarray}
\nonumber&&\int_{\infty}^{\fz}\frac{1}{\left(1-\sqrtz\right)^2}d\fz^{-2}\\
&=&\frac{\fz^{-2}}{4\lagb\left(1-\sqrtz\right)}+\frac{1-4\lagb}{32\lagb^2}\lncom,
\end{eqnarray}
\begin{eqnarray}
\nonumber&&\int_{\infty}^{\fz}\frac{\sqrt{1-4\lagb}}{\sqrtz\left(\sqrtz+\sqrt{1-4\lagb}\right)}d\fz^{-2}\\
\nonumber&=&\int_{\infty}^{\fz}\left(\frac{1}{\sqrtz}-\frac{1}{\sqrtz+\sqrt{1-4\lagb}}\right)d\fz^{-2}\\
&=&\frac{\fz^{-2}}{\sqrtz+\sqrt{1-4\lagb}},
\end{eqnarray}
\begin{eqnarray}
\nonumber&&\int_{\infty}^{\fz}\frac{1}{\sqrtz\left(1-\sqrtz\right)}d\fz^{-2}\\
\nonumber&=&\int_{\infty}^{\fz}\left(\frac{1}{\sqrtz}+\frac{1}{1-\sqrtz}\right)d\fz^{-2}\\
&=&\frac{1}{8\lagb}\lncom,
\end{eqnarray}
\begin{eqnarray}
\nonumber&&\int_{\infty}^{\fz}\frac{1+\sqrt{1-4\lagb}}{\left(1-\sqrtz\right)\left(\sqrtz+\sqrt{1-4\lagb}\right)}d\fz^{-2}\\
\nonumber&=&\int_{\infty}^{\fz}\left(\frac{1}{1-\sqrtz}+\frac{1}{\sqrtz+\sqrt{1-4\lagb}}\right)d\fz^{-2}\\
&=&-\frac{\fz^{-2}}{\sqrtz+\sqrt{1-4\lagb}}+\frac{1}{8\lagb}\lncom,
\end{eqnarray}
\begin{eqnarray}
\nonumber&&\int_{\infty}^{\fz}\frac{1+\sqrt{1-4\lagb}}{\sqrtz\left(1-\sqrtz\right)\left(\sqrtz+\sqrt{1-4\lagb}\right)}d\fz^{-2}\\
\nonumber&=&\int_{\infty}^{\fz}\Biggl[\left(1+\frac{1}{\sqrt{1-4\lagb}}\right)\frac{1}{\sqrtz}+\frac{1}{1-\sqrtz}\\
\nonumber&&-\frac{1}{\sqrt{1-4\lagb}}\frac{1}{\sqrtz+\sqrt{1-4\lagb}}\Biggr]d\fz^{-2}\\
&=&\frac{\fz^{-2}}{\sqrt{1-4\lagb}\left(\sqrtz+\sqrt{1-4\lagb}\right)}+\frac{1}{8\lagb}\lncom.
\end{eqnarray}
\begin{comment}%comment20181225
The following one may be useless.
\begin{eqnarray}
\nonumber&&\int_{\infty}^{\fz}\frac{(1-4\lagb)\fz^4}{\sqrtz\left(1-\sqrtz\right)}d\fz^{-2}\\
\nonumber&=&\int_{\infty}^{\fz}\Biggl[\frac{-4\lagb}{\sqrtz}+\frac{1-4\lagb}{1-\sqrtz}+\frac{1-\sqrt{1-4\lagb}}{2\left(\sqrtz+\sqrt{1-4\lagb}\right)}\\
\nonumber&&+\frac{1+\sqrt{1-4\lagb}}{2\left(\sqrtz-\sqrt{1-4\lagb}\right)}\Biggr]d\fz^{-2}\\
\nonumber&=&\int_{\infty}^{\fz}\Biggl[\frac{1-4\lagb}{\sqrtz\left(1-\sqrtz\right)}+\frac{\sqrt{1-4\lagb}}{2\lagb}\fz^4\\
\nonumber&&-\frac{\sqrt{1-4\lagb}}{\sqrtz\left(\sqrtz+\sqrt{1-4\lagb}\right)}\Biggr]d\fz^{-2}\\
\nonumber&=&\int_{\infty}^{\fz}d\Biggl[\frac{1-4\lagb}{8\lagb}\lncom-\frac{\sqrt{1-4\lagb}}{2\lagb}\fz^2\\
\nonumber&&-\frac{\fz^{-2}}{\sqrtz+\sqrt{1-4\lagb}}\Biggr]\\
\end{eqnarray}
\begin{equation}
\lim_{\fz\rightarrow1}\ln\frac{(1-4\lagb)\left(1-\sqrtz\right)^2\left(1+\fz^{-2}\right)\left(\sqrtz+\fz^{-2}\right)}{(8\lagb)^2\left(1-\fz^{-2}\right)\left(\sqrtz-\fz^{-2}\right)}=\mathcal{O}\left(1-\fz^{-2}\right)
\end{equation}

The following three are useful.
\end{comment}%comment20181225

\begin{comment}
\begin{eqnarray}
\nonumber&&\lim_{\fz\rightarrow1}\frac{C_0\tq^2(1-4\lagb)}{128\lagb\left(1+\sqrt{1-4\lagb}\right)}\left(1-\sqrtz\right)\lncom\\
\nonumber&=&\lim_{\fz\rightarrow1}\frac{C_0\tq^2(1-4\lagb)}{128\lagb\left(1+\sqrt{1-4\lagb}\right)}2\lagb\left(1-\fz^{-4}\right)\ln\frac{\left(1+\fz^{-2}\right)^2\left(\sqrtz+\fz^{-2}\right)^2}{(1-4\lagb)\left(1-\fz^{-4}\right)^2}\\
\nonumber&=&-\frac{C_0\tq^2(1-4\lagb)}{32\left(1+\sqrt{1-4\lagb}\right)}\lim_{\fz\rightarrow1}\left(1-\fz^{-4}\right)\ln\left(1-\fz^{-4}\right)\\
&=&0,
\end{eqnarray}
by L'hospital's rule, the last line of Eq. \eqref{B2} is
\begin{eqnarray}
\nonumber&&\lim_{\fz\rightarrow1}\frac{C_0\tq^2(1-4\lagb)}{128\lagb\left(1+\sqrt{1-4\lagb}\right)}\left(1-\sqrtz\right)\lncom\\
%\nonumber&=&-\frac{C_0\tq^2(1-4\lagb)}{128\lagb\left(1+\sqrt{1-4\lagb}\right)}\lim_{\fz\rightarrow1}\frac{\ln\left(\fz^2-1\right)+\ln\left[\sqrt{(1-4\lagb)\fz^4+4\lagb}-1\right]}{\left(1-\sqrtz\right)^{-1}}\\
\nonumber&=&-\frac{C_0\tq^2(1-4\lagb)}{128\lagb\left(1+\sqrt{1-4\lagb}\right)}\lim_{\fz\rightarrow1}\frac{\frac{2\fz}{\fz^2-1}+\left[\sqrt{(1-4\lagb)\fz^4+4\lagb}-1\right]^{-1}\times\frac{4(1-4\lagb)\fz^3}{2\sqrt{(1-4\lagb)\fz^4+4\lagb}}}{\left(1-\sqrtz\right)^{-2}\times\frac{-16\lagb\fz^{-5}}{2\sqrtz}}\\
%\nonumber&=&-\frac{C_0\tq^2(1-4\lagb)}{128\lagb\left(1+\sqrt{1-4\lagb}\right)}\lim_{\fz\rightarrow1}\frac{\frac{2}{\fz^2-1}+\left[(1-2\lagb)(\fz^4-1)\right]^{-1}\times2(1-4\lagb)}{\left[2\lagb\left(1-\fz^{-4}\right)\right]^{-2}\times(-8\lagb)}\\
&=&0,
\end{eqnarray}
\end{comment}

In terms of $f$ defined in Eqs. \eqref{fGBz}, it is not hard to show
\begin{eqnarray}\label{limln0}
\nonumber&&\lncom\\
\nonumber&=&\ln\frac{\left(1-\sqrtz\right)^2\left(1+\fz^{-2}\right)\left(\sqrtz+\fz^{-2}\right)}{4\lagb^2\left(1-\fz^{-2}\right)\left(\sqrtz-\fz^{-2}\right)}-2\ln f\\
\nonumber&=&\ln\frac{4\left(1+\fz^{-2}\right)^2\left(\sqrtz+\fz^{-2}\right)^2}{(1-4\lagb)\left(1+\sqrtz\right)^2}-2\ln f\\
%\nonumber&=&\ln\frac{\left(1+\fz^{-2}\right)^2\left(\sqrtz+\fz^{-2}\right)^2}{4\left(1+\sqrtz\right)^2}+\ln\frac{16}{1-4\lagb}-2\ln f\\
&=&2\ln\frac{\left(1+\fz^{-2}\right)\left(\sqrtz+\fz^{-2}\right)}{2\left(1+\sqrtz\right)}+\ln\frac{16}{1-4\lagb}-2\ln f,
\end{eqnarray}

All of the above formulas in this appendix will be utilized frequently and implicitly in subsequent appendices.

Sometimes, it is also necessary to take limits of the following expressions.
\begin{eqnarray}\label{limln1}
\nonumber&&\fz\left.\lncom\right|_{\fz=\infty}\\
\nonumber&=&2\fz\left.\left(\fz^{-2}+\frac{2\lagb}{1-4\lagb}\fz^{-4}+\frac{\fz^{-2}}{\sqrt{1-4\lagb}}\right)\right|_{\fz=\infty}\\
&=&0,
\end{eqnarray}
\begin{eqnarray}\label{limln2}
\nonumber&&\left(1-\sqrtz\right)\left.\lncom\right|_{\fz=1}\\
\nonumber&=&\frac{4\lagb\left(1-\fz^{-4}\right)}{1+\sqrtz}\left.\ln\frac{\left(1+\fz^{-2}\right)^2\left(\sqrtz+\fz^{-2}\right)^2}{(1-4\lagb)\left(1-\fz^{-4}\right)^2}\right|_{\fz=1}\\
%\nonumber&=&2\lagb\left(1-\fz^{-2}\right)\left.\ln\frac{16}{\left(1-4\lagb\right)\left(1-\fz^{-4}\right)^2}\right|_{\fz=1}\\
\nonumber&=&-4\lagb\left(1-\fz^{-4}\right)\left.\ln\left(1-\fz^{-4}\right)\right|_{\fz=1}\\
&=&0,
\end{eqnarray}
\begin{eqnarray}\label{limln3}
\nonumber&&\left.\frac{1}{1-\sqrtz}\ln\frac{\left(1+\fz^{-2}\right)\left(\sqrtz+\fz^{-2}\right)}{2\left(1+\sqrtz\right)}\right|_{\fz=1}\\
\nonumber&=&\left.\frac{1+\sqrtz}{4\lagb\left(1-\fz^{-4}\right)}\ln\left[1+\frac{\left(\fz^{-2}-1\right)\left(\sqrtz+\fz^{-2}+2\right)}{2\left(1+\sqrtz\right)}\right]\right|_{\fz=1}\\
\nonumber&=&\left.\frac{1+\sqrtz}{4\lagb\left(1-\fz^{-4}\right)}\times\frac{\left(\fz^{-2}-1\right)\left(\sqrtz+\fz^{-2}+2\right)}{2\left(1+\sqrtz\right)}\right|_{\fz=1}\\
%\nonumber&=&\left.\frac{1}{2\lagb\left(1-\fz^{-4}\right)}\times\left(\fz^{-2}-1\right)\right|_{\fz=1}\\
&=&-\frac{1}{4\lagb}
\end{eqnarray}
and likewise
\begin{eqnarray}\label{limln4}
\nonumber&&\left.\left(1-\sqrtz\right)(\ln f)\right|_{\fz=1}\\
\nonumber&=&\left.2\lagb\left(1-\fz^{-4}\right)(\ln f)\right|_{\fz=1}\\
\nonumber&=&\left.2\lagb\left(1-\fz^{-4}\right)\ln\left(1-\fz^{-4}\right)\right|_{\fz=1}\\
&=&0,
\end{eqnarray}
\begin{eqnarray}\label{limln5}
\nonumber&&\left(1-\sqrtz\right)(\ln f)\left.\lncom\right|_{\fz=1}\\
\nonumber&=&2\lagb\left(1-\fz^{-4}\right)\left.\left[\ln\left(1-\fz^{-4}\right)\right]^2\right|_{\fz=1}\\
&=&0.
\end{eqnarray}

\section{Working out integrations \eqref{dg1} and \eqref{g1}}\label{app-g1}
Corresponding to Eqs. \eqref{dg1}, \eqref{g1}, there are two integrations that should be and can be worked out in this appendix. The first one is
\begin{eqnarray}\label{B1}
\nonumber&&\fz^5\sqrtz\partial_{\fz}\left[g_1-\left(\frac{i\tom_1}{4\Nsh}\ln f\right)g_0\right]\\
\nonumber&=&\int_{\infty}^{\fz}\frac{16C_0\Nsh^{-2}\tom_1^2\lagb^2\left(2-\sqrtz\right)}{\tq^2(1-4\lagb)\fz^5\left(1-\sqrtz\right)^2}d\fz\\
\nonumber&&-\int_{\infty}^{\fz}\frac{C_0\tq^2(1-4\lagb)\fz\left(\sqrtz-\sqrt{1-4\lagb}\right)}{4\sqrtz\left(1-\sqrtz\right)}d\fz+C_1\\
\nonumber&=&-\int_{\infty}^{\fz}\frac{2C_0\Nsh^{-2}\tom_1^2\lagb\sqrtz\left(2-\sqrtz\right)}{\tq^2(1-4\lagb)\left(1-\sqrtz\right)^2}d\sqrtz\\
\nonumber&&-\int_{\infty}^{\fz}\frac{C_0\tq^2(1-4\lagb)\fz}{4\sqrtz\left(1-\sqrtz\right)}\frac{4\lagb\fz^{-4}}{\sqrtz+\sqrt{1-4\lagb}}d\fz\\
\nonumber&&+C_1\\
\nonumber&=&\int_{\infty}^{\fz}\frac{2C_0\Nsh^{-2}\tom_1^2\lagb}{\tq^2(1-4\lagb)}d\left(\sqrtz-\frac{1}{1-\sqrtz}\right)\\
\nonumber&&+\int_{\infty}^{\fz}\frac{C_0\tq^2\lagb(1-4\lagb)}{2\sqrtz\left(1-\sqrtz\right)\left(\sqrtz+\sqrt{1-4\lagb}\right)}d\fz^{-2}\\
\nonumber&&+C_1\\
\nonumber&=&\frac{2C_0\Nsh^{-2}\tom_1^2\lagb}{\tq^2(1-4\lagb)}\Biggl(\sqrtz-\frac{1}{1-\sqrtz}-\sqrt{1-4\lagb}\\
\nonumber&&+\frac{1}{1-\sqrt{1-4\lagb}}\Biggr)+\frac{C_0\tq^2\lagb(1-4\lagb)}{2\left(1+\sqrt{1-4\lagb}\right)}\Biggl[\frac{\fz^{-2}}{\sqrt{1-4\lagb}\left(\sqrtz+\sqrt{1-4\lagb}\right)}\\
&&+\frac{1}{8\lagb}\lncom\Biggr]+C_1.
\end{eqnarray}
The second one is
\begin{eqnarray}\label{B2}
\nonumber&&g_1-\left(\frac{i\tom_1}{4\Nsh}\ln f\right)g_0\\
\nonumber&=&\int_{\infty}^{\fz}\frac{1}{\fz^5\sqrtz}\times\eqref{B1}d\fz\\
\nonumber&=&-\frac{1}{8\lagb}\int_{\infty}^{\fz}\Biggl[\frac{2C_0\Nsh^{-2}\tom_1^2\lagb}{\tq^2(1-4\lagb)}\Biggl(\sqrtz-\frac{1}{1-\sqrtz}-\sqrt{1-4\lagb}\\
\nonumber&&+\frac{1}{1-\sqrt{1-4\lagb}}\Biggr)+C_1\Biggr]d\sqrtz\\
\nonumber&&+\int_{\infty}^{\fz}\frac{1}{\fz^5\sqrtz}\left[\frac{C_0\tq^2\lagb\sqrt{1-4\lagb}}{2\left(1+\sqrt{1-4\lagb}\right)}\frac{\fz^{-2}}{\sqrtz+\sqrt{1-4\lagb}}\right]d\fz\\
\nonumber&&-\frac{1}{8\lagb}\int_{\infty}^{\fz}\left[\frac{C_0\tq^2(1-4\lagb)}{16\left(1+\sqrt{1-4\lagb}\right)}\lncom\right]d\sqrtz\\
\nonumber&=&-\int_{\infty}^{\fz}\frac{C_0\Nsh^{-2}\tom_1^2}{4\tq^2(1-4\lagb)}d\Biggl[\frac{1}{2}\left(1-4\lagb+4\lagb\fz^{-4}\right)+\ln\left(1-\sqrtz\right)\\
\nonumber&&+\left(\frac{1}{1-\sqrt{1-4\lagb}}-\sqrt{1-4\lagb}\right)\sqrtz\Biggr]-\frac{C_1}{8\lagb}\int_{\infty}^{\fz}d\sqrtz\\
\nonumber&&-\frac{C_0\tq^2\lagb\sqrt{1-4\lagb}}{4\left(1+\sqrt{1-4\lagb}\right)}\int_{\infty}^{\fz}\frac{\fz^{-4}}{\sqrtz\left(\sqrtz+\sqrt{1-4\lagb}\right)}d\fz^{-2}\\
\nonumber&&-\frac{C_0\tq^2(1-4\lagb)}{128\lagb\left(1+\sqrt{1-4\lagb}\right)}\sqrtz\lncom\\
\nonumber&&+\frac{1}{8\lagb}\int_{\infty}^{\fz}\left[\frac{C_0\tq^2(1-4\lagb)}{16\left(1+\sqrt{1-4\lagb}\right)}\sqrtz\right]d\lncom\\
\nonumber&=&-\frac{C_0\Nsh^{-2}\tom_1^2}{4\tq^2(1-4\lagb)}\Biggl[2\lagb\fz^{-4}+\ln\frac{1-\sqrtz}{1-\sqrt{1-4\lagb}}+\left(\frac{1}{1-\sqrt{1-4\lagb}}-\sqrt{1-4\lagb}\right)\\
\nonumber&&\times\left(\sqrtz-\sqrt{1-4\lagb}\right)\Biggr]-\frac{C_1}{8\lagb}\left(\sqrtz-\sqrt{1-4\lagb}\right)\\
\nonumber&&-\frac{C_0\tq^2\sqrt{1-4\lagb}}{16\left(1+\sqrt{1-4\lagb}\right)}\int_{\infty}^{\fz}\left(1-\frac{\sqrt{1-4\lagb}}{\sqrtz}\right)d\fz^{-2}\\
\nonumber&&-\frac{C_0\tq^2(1-4\lagb)}{128\lagb\left(1+\sqrt{1-4\lagb}\right)}\sqrtz\lncom\\
\nonumber&&+\frac{C_0\tq^2(1-4\lagb)}{16\left(1+\sqrt{1-4\lagb}\right)}\int_{\infty}^{\fz}\frac{1}{1-\sqrtz}d\fz^{-2}\\
\nonumber&=&-\frac{C_0\Nsh^{-2}\tom_1^2}{4\tq^2(1-4\lagb)}\Biggl[2\lagb\fz^{-4}+\ln\frac{1-\sqrtz}{1-\sqrt{1-4\lagb}}+\left(\frac{1}{1-\sqrt{1-4\lagb}}-\sqrt{1-4\lagb}\right)\\
\nonumber&&\times\left(\sqrtz-\sqrt{1-4\lagb}\right)\Biggr]-\frac{C_1}{8\lagb}\left(\sqrtz-\sqrt{1-4\lagb}\right)\\
\nonumber&&-\frac{C_0\tq^2\sqrt{1-4\lagb}}{16\left(1+\sqrt{1-4\lagb}\right)}\fz^{-2}+\frac{C_0\tq^2(1-4\lagb)}{128\lagb\left(1+\sqrt{1-4\lagb}\right)}\left(1-\sqrtz\right)\\
&&\times\lncom.
\end{eqnarray}

\section{Divergent terms of \eqref{g2v}}\label{app-g2v}
In Eqs. \eqref{dg2}, \eqref{g2}, we defined two other integrations. Although we can work them out as we have done in the previous appendix, the details are rather cumbersome. However, in the present paper, what really matter are the divergent terms of \eqref{g2v}. Thus we will take a short cut that is easier for the readers to follow.

Firstly, we note that
\begin{eqnarray}\label{C1}
\nonumber&&\fz^5\sqrtz\partial_{\fz}\left[g_2-\left(\frac{i\tom_1}{4\Nsh}\ln f\right)g_1-\left(\frac{i\tom_2}{4\Nsh}\ln f\right)g_0-\frac{1}{2}\left(\frac{\tom_1}{4\Nsh}\ln f\right)^2g_0\right]\\
\nonumber&=&\int_{\infty}^{\fz}\frac{16\Nsh^{-2}\tom_1^2\lagb^2\left(2-\sqrtz\right)}{\tq^2(1-4\lagb)\fz^5\left(1-\sqrtz\right)^2}\times\eqref{B1}d\fz\\
\nonumber&&+\int_{\infty}^{\fz}\frac{2\tq^2\lagb(1-4\lagb)\fz}{\sqrtz\left(1-\sqrtz\right)}\times\eqref{B2}d\fz\\
\nonumber&&+\int_{\infty}^{\fz}\frac{32C_0\Nsh^{-2}\tom_1\tom_2\lagb^2\left(2-\sqrtz\right)}{\tq^2(1-4\lagb)\fz^5\left(1-\sqrtz\right)^2}d\fz\\
\nonumber&&+\int_{\infty}^{\fz}\frac{32C_0\Nsh^{-4}\tom_1^4\lagb^3\left(1-4\lagb+4\lagb\fz^{-4}\right)\left(2-\sqrtz\right)}{\tq^4(1-4\lagb)^2\fz^5\left(1-\sqrtz\right)^3}d\fz\\
\nonumber&&+\int_{\infty}^{\fz}\frac{C_0\Nsh^{-2}\tom_1^2\lagb\fz\sqrtz}{2\left(1-\sqrtz\right)^2}\left(\sqrtz-\sqrt{1-4\lagb}\right)d\fz+C_2\\
\nonumber&=&\int_{\infty}^{\fz}\frac{16\Nsh^{-2}\tom_1^2\lagb^2\left(2-\sqrtz\right)}{\tq^2(1-4\lagb)\fz^5\left(1-\sqrtz\right)^2}\Biggl\{\frac{2C_0\Nsh^{-2}\tom_1^2\lagb}{\tq^2(1-4\lagb)}\Biggl(-1-\sqrt{1-4\lagb}\\
\nonumber&&+\frac{1}{1-\sqrt{1-4\lagb}}\Biggr)+\frac{C_0\tq^2\lagb(1-4\lagb)}{2\left(1+\sqrt{1-4\lagb}\right)}\Biggl[\frac{\fz^{-2}}{\sqrt{1-4\lagb}\left(\sqrtz+\sqrt{1-4\lagb}\right)}\\
\nonumber&&+\frac{1}{8\lagb}\lncom\Biggr]+C_1\Biggr\}d\fz\\
\nonumber&&+\int_{\infty}^{\fz}\frac{2\tq^2\lagb(1-4\lagb)\fz}{\sqrtz\left(1-\sqrtz\right)}\times\eqref{B2}d\fz\\
\nonumber&&+\int_{\infty}^{\fz}\frac{32C_0\Nsh^{-2}\tom_1\tom_2\lagb^2\left(2-\sqrtz\right)}{\tq^2(1-4\lagb)\fz^5\left(1-\sqrtz\right)^2}d\fz\\
\nonumber&&+\int_{\infty}^{\fz}\frac{C_0\Nsh^{-2}\tom_1^2\lagb\fz\sqrtz}{2\left(1-\sqrtz\right)^2}\left(\sqrtz-\sqrt{1-4\lagb}\right)d\fz+C_2.\\
\end{eqnarray}
It is not hard to check that $\left.\eqref{C1}\right|_{\fz=\infty}=0$. This has been used in the last step of Eq. \eqref{g2}. Secondly, we make the transformation
\begin{eqnarray}\label{C2}
\nonumber&&\int_{\infty}^{\fz}\left(1-\sqrtz\right)\partial_{\fz}\eqref{C1}d\fz\\
\nonumber&=&\int_{\infty}^{\fz}\frac{16\Nsh^{-2}\tom_1^2\lagb^2\left(2-\sqrtz\right)}{\tq^2(1-4\lagb)\fz^5\left(1-\sqrtz\right)}\times\eqref{B1}d\fz\\
\nonumber&&+\int_{\infty}^{\fz}\frac{2\tq^2\lagb(1-4\lagb)\fz}{\sqrtz}\times\eqref{B2}d\fz+\int_{\infty}^{\fz}\frac{32C_0\Nsh^{-2}\tom_1\tom_2\lagb^2\left(2-\sqrtz\right)}{\tq^2(1-4\lagb)\fz^5\left(1-\sqrtz\right)}d\fz\\
\nonumber&&+\int_{\infty}^{\fz}\frac{32C_0\Nsh^{-4}\tom_1^4\lagb^3\left(1-4\lagb+4\lagb\fz^{-4}\right)\left(2-\sqrtz\right)}{\tq^4(1-4\lagb)^2\fz^5\left(1-\sqrtz\right)^2}d\fz\\
\nonumber&&+\int_{\infty}^{\fz}\frac{C_0\Nsh^{-2}\tom_1^2\lagb\fz\sqrtz}{2\left(1-\sqrtz\right)}\left(\sqrtz-\sqrt{1-4\lagb}\right)d\fz\\
\nonumber&=&\int_{\infty}^{\fz}\frac{16\Nsh^{-2}\tom_1^2\lagb^2\left(2-\sqrtz\right)}{\tq^2(1-4\lagb)\fz^5\left(1-\sqrtz\right)}\Biggl\{\frac{2C_0\Nsh^{-2}\tom_1^2\lagb}{\tq^2(1-4\lagb)}\Biggl(-1-\sqrt{1-4\lagb}\\
\nonumber&&+\frac{1}{1-\sqrt{1-4\lagb}}\Biggr)+\frac{C_0\tq^2\lagb(1-4\lagb)}{2\left(1+\sqrt{1-4\lagb}\right)}\Biggl[\frac{\fz^{-2}}{\sqrt{1-4\lagb}\left(\sqrtz+\sqrt{1-4\lagb}\right)}\\
\nonumber&&+\frac{1}{8\lagb}\lncom\Biggr]+C_1\Biggr\}d\fz\\
\nonumber&&+\int_{\infty}^{\fz}\frac{2\tq^2\lagb(1-4\lagb)\fz}{\sqrtz}\times\eqref{B2}d\fz+\int_{\infty}^{\fz}\frac{32C_0\Nsh^{-2}\tom_1\tom_2\lagb^2\left(2-\sqrtz\right)}{\tq^2(1-4\lagb)\fz^5\left(1-\sqrtz\right)}d\fz\\
\nonumber&&+\int_{\infty}^{\fz}\frac{C_0\Nsh^{-2}\tom_1^2\lagb\fz\sqrtz}{2\left(1-\sqrtz\right)}\left(\sqrtz-\sqrt{1-4\lagb}\right)d\fz\\
\nonumber&=&-\int_{\infty}^{\fz}\frac{2\Nsh^{-2}\tom_1^2\lagb\left(2-\sqrtz\right)}{\tq^2(1-4\lagb)\left(1-\sqrtz\right)}\Biggl[\frac{2C_0\Nsh^{-2}\tom_1^2\lagb}{\tq^2\left(1-\sqrt{1-4\lagb}\right)}+\frac{C_0\tq^2(1-4\lagb)}{16\left(1+\sqrt{1-4\lagb}\right)}\\
\nonumber&&\times\lncom+C_1\Biggr]\sqrtz d\sqrtz\\
\nonumber&&+\frac{8C_0\Nsh^{-2}\tom_1^2\lagb^3}{\sqrt{1-4\lagb}\left(1+\sqrt{1-4\lagb}\right)}\int_{\infty}^{\fz}\frac{2-\sqrtz}{1-\sqrtz}\frac{\sqrtz-\sqrt{1-4\lagb}}{-8\lagb}d\fz^{-2}\\
\nonumber&&+\int_{\infty}^{\fz}\frac{2\tq^2\lagb(1-4\lagb)\fz}{\sqrtz}\times\eqref{B2}d\fz\\
\nonumber&&-\int_{\infty}^{\fz}\frac{4C_0\Nsh^{-2}\tom_1\tom_2\lagb}{\tq^2(1-4\lagb)}d\left[2\lagb\fz^{-4}-\sqrtz-\ln\left(1-\sqrtz\right)\right]\\
\nonumber&&-\frac{C_0\Nsh^{-2}\tom_1^2\lagb^2}{1+\sqrt{1-4\lagb}}\int_{\infty}^{\fz}\Biggl[\frac{1}{1-\sqrtz}-\frac{\sqrt{1-4\lagb}}{\sqrtz+\sqrt{1-4\lagb}}\Biggr]d\fz^{-2}.\\
\end{eqnarray}
For later use, we can also prove that
\begin{eqnarray}\label{C2div}
\nonumber&&-\int_{\infty}^{\fz}\frac{2\tq^2\lagb(1-4\lagb)\fz}{\sqrtz}\frac{C_0\Nsh^{-2}\tom_1^2}{4\tq^2(1-4\lagb)}\ln\frac{1-\sqrtz}{1-\sqrt{1-4\lagb}}d\fz\\
\nonumber&=&-\int_{\infty}^{\fz}\frac{C_0\Nsh^{-2}\tom_1^2\lagb}{4(1-4\lagb)}\ln\frac{1-\sqrtz}{1-\sqrt{1-4\lagb}}d\sqrt{\fz^4-4\lagb\fz^4+4\lagb}\\
\nonumber&=&\int_{\infty}^{\fz}\frac{C_0\Nsh^{-2}\tom_1^2\lagb}{4(1-4\lagb)}\sqrt{\fz^4-4\lagb\fz^4+4\lagb}d\ln\frac{1-\sqrtz}{1-\sqrt{1-4\lagb}}\\
\nonumber&&-\frac{C_0\Nsh^{-2}\tom_1^2\lagb}{4(1-4\lagb)}\sqrt{\fz^4-4\lagb\fz^4+4\lagb}\ln\frac{1-\sqrtz}{1-\sqrt{1-4\lagb}}\\
\nonumber&&+\left.\frac{C_0\Nsh^{-2}\tom_1^2\lagb}{4(1-4\lagb)}\sqrt{\fz^4-4\lagb\fz^4+4\lagb}\ln\frac{1-\sqrtz}{1-\sqrt{1-4\lagb}}\right|_{\fz=\infty}\\
\nonumber&=&-\frac{C_0\Nsh^{-2}\tom_1^2\lagb^2}{1-4\lagb}\int_{\infty}^{\fz}\frac{1}{1-\sqrtz}d\fz^{-2}\\
\nonumber&&-\frac{C_0\Nsh^{-2}\tom_1^2\lagb}{4(1-4\lagb)}\sqrt{\fz^4-4\lagb\fz^4+4\lagb}\ln\frac{1-\sqrtz}{1-\sqrt{1-4\lagb}}\\
&&+\left.\frac{C_0\Nsh^{-2}\tom_1^2\lagb}{4(1-4\lagb)}\sqrt{\fz^4-4\lagb\fz^4+4\lagb}\times\frac{-2\lagb\fz^{-4}}{\sqrt{1-4\lagb}\left(1-\sqrt{1-4\lagb}\right)}\right|_{\fz=\infty}.
\end{eqnarray}

Here we care about the terms that are divergent at $\fz=1$, so we will use $\simeq$ to denote equivalence up to convergent terms. By definition $g_0(1)$, $g_1(1)$, $g_2(1)$ are finite, while $\ln f(1)$ is divergent. Therefore, on the left hand side of Eq. \eqref{g2v}, the divergent terms are
\begin{eqnarray}\label{g2vleft}
%\nonumber&&-8\lagb\left.\left[-\left(\frac{i\tom_1}{4\Nsh}\ln f\right)g_1-\left(\frac{i\tom_2}{4\Nsh}\ln f\right)g_0-\frac{1}{2}\left(\frac{\tom_1}{4\Nsh}\ln f\right)^2g_0\right]\right|_{\fz=1}=\\
\nonumber&&-8\lagb\left.\left\{-\left(\frac{i\tom_1}{4\Nsh}\ln f\right)\left[g_1-\left(\frac{i\tom_1}{4\Nsh}\ln f\right)g_0\right]-\left(\frac{i\tom_2}{4\Nsh}\ln f\right)g_0+\frac{1}{2}\left(\frac{\tom_1}{4\Nsh}\ln f\right)^2g_0\right\}\right|_{\fz=1}\\
\nonumber&=&\frac{2i\tom_1\lagb}{\Nsh}\left.\left[(\ln f)\times\eqref{B2}\right]\right|_{\fz=1}-\frac{iC_0\tom_2}{4\Nsh}\left(1-\sqrt{1-4\lagb}\right)\left.(\ln f)\right|_{\fz=1}+\frac{C_0\tom_1^2}{32\Nsh^2}\left(1-\sqrt{1-4\lagb}\right)\left.(\ln f)^2\right|_{\fz=1}\\
\nonumber&=&\frac{2i\tom_1\lagb}{\Nsh}\left.(\ln f)\right|_{\fz=1}\Biggl\{-\frac{C_0\Nsh^{-2}\tom_1^2}{4\tq^2(1-4\lagb)}\left[\left.(\ln f)\right|_{\fz=1}-\ln\frac{1-\sqrt{1-4\lagb}}{2\lagb}+1+\frac{1}{2}\left(1-\sqrt{1-4\lagb}\right)^2\right]\\
\nonumber&&-\frac{C_1}{8\lagb}\left(1-\sqrt{1-4\lagb}\right)-\frac{C_0\tq^2\sqrt{1-4\lagb}}{16\left(1+\sqrt{1-4\lagb}\right)}\Biggr\}\\
&&-\frac{iC_0\tom_2}{4\Nsh}\left(1-\sqrt{1-4\lagb}\right)\left.(\ln f)\right|_{\fz=1}+\frac{C_0\tom_1^2}{32\Nsh^2}\left(1-\sqrt{1-4\lagb}\right)\left.(\ln f)^2\right|_{\fz=1},
\end{eqnarray}
in which we have made use of Eqs. \eqref{g0v}, \eqref{limln3}, \eqref{limln4}. By L'hospital's rule, the first term on the right hand side of Eq. \eqref{g2v} tends to
\begin{eqnarray}
\nonumber&&-\left.\left(1-\sqrtz\right)\times\eqref{C1}\right|_{\fz=1}\\
&=&\left.\frac{\sqrtz\left(1-\sqrtz\right)^2}{8\lagb\fz^3}\times\partial_{\fz}\eqref{C1}\right|_{\fz=1}.
\end{eqnarray}
Making use of Eqs. \eqref{limln0}, \eqref{limln2}, \eqref{limln4}, we find it has a divergent term
\begin{eqnarray}\label{g2vrighta}
\nonumber&&-\left.\left(1-\sqrtz\right)\times\eqref{C1}\right|_{\fz=1}\\
\nonumber&\simeq&\frac{\sqrtz\left(1-\sqrtz\right)^2}{8\lagb\fz^3}\frac{16\Nsh^{-2}\tom_1^2\lagb^2\left(2-\sqrtz\right)}{\tq^2(1-4\lagb)\fz^5\left(1-\sqrtz\right)^2}\\
\nonumber&&\times\left.\frac{C_0\tq^2(1-4\lagb)}{16\left(1+\sqrt{1-4\lagb}\right)}\lncom\right|_{\fz=1}\\
%\nonumber&\simeq&\left.\frac{\sqrtz}{8\lagb\fz^3}\frac{16\Nsh^{-2}\tom_1^2\lagb^2\left(2-\sqrtz\right)}{\tq^2(1-4\lagb)\fz^5}\frac{C_0\tq^2(1-4\lagb)}{16\left(1+\sqrt{1-4\lagb}\right)}(-2\ln f)\right|_{\fz=1}\\
&\simeq&-\frac{C_0\Nsh^{-2}\tom_1^2\lagb}{4\left(1+\sqrt{1-4\lagb}\right)}\left.(\ln f)\right|_{\fz=1}.
\end{eqnarray}
Further more, making use of Eqs. \eqref{limln0}, \eqref{limln1}, \eqref{limln2}, \eqref{limln3}, \eqref{C2div}, we can get the remainder divergent terms on the right hand side,
\begin{eqnarray}\label{g2vrightb}
\nonumber&&\left.\eqref{C2}\right|_{\fz=1}\\
%\nonumber&=&-\int_{\infty}^1\frac{2\Nsh^{-2}\tom_1^2\lagb\left(2-\sqrtz\right)}{\tq^2(1-4\lagb)\left(1-\sqrtz\right)}\Biggl[\frac{2C_0\Nsh^{-2}\tom_1^2\lagb}{\tq^2\left(1-\sqrt{1-4\lagb}\right)}+\frac{C_0\tq^2(1-4\lagb)}{16\left(1+\sqrt{1-4\lagb}\right)}\\
%\nonumber&&\times\lncom+C_1\Biggr]\sqrtz d\sqrtz\\
%\nonumber&&+\frac{8C_0\Nsh^{-2}\tom_1^2\lagb^3}{\sqrt{1-4\lagb}\left(1+\sqrt{1-4\lagb}\right)}\int_{\infty}^1\frac{2-\sqrtz}{1-\sqrtz}\frac{\sqrtz-\sqrt{1-4\lagb}}{-8\lagb}d\fz^{-2}\\
%\nonumber&&+\int_{\infty}^1\frac{2\tq^2\lagb(1-4\lagb)\fz}{\sqrtz}\times\eqref{B2}d\fz\\
%\nonumber&&-\int_{\infty}^1\frac{4C_0\Nsh^{-2}\tom_1\tom_2\lagb}{\tq^2(1-4\lagb)}d\left[2\lagb\fz^{-4}-\sqrtz-\ln\left(1-\sqrtz\right)\right]\\
%\nonumber&&-\frac{C_0\Nsh^{-2}\tom_1^2\lagb^2}{1+\sqrt{1-4\lagb}}\int_{\infty}^1\Biggl[\frac{1}{1-\sqrtz}-\frac{\sqrt{1-4\lagb}}{\sqrtz+\sqrt{1-4\lagb}}\Biggr]d\fz^{-2}\\
\nonumber&\simeq&-\int_{\infty}^1\frac{2\Nsh^{-2}\tom_1^2\lagb}{\tq^2(1-4\lagb)\left(1-\sqrtz\right)}\Biggl\{\frac{2C_0\Nsh^{-2}\tom_1^2\lagb}{\tq^2\left(1-\sqrt{1-4\lagb}\right)}+C_1+\frac{C_0\tq^2(1-4\lagb)}{16\left(1+\sqrt{1-4\lagb}\right)}\\
\nonumber&&\times\left[2\ln\frac{\left(1+\fz^{-2}\right)\left(\sqrtz+\fz^{-2}\right)}{2\left(1+\sqrtz\right)}+\ln\frac{16}{1-4\lagb}-2\ln f\right]\Biggr\}d\sqrtz\\
\nonumber&&-\frac{C_0\Nsh^{-2}\tom_1^2\lagb^2}{\sqrt{1-4\lagb}\left(1+\sqrt{1-4\lagb}\right)}\int_{\infty}^1\frac{1-\sqrt{1-4\lagb}}{1-\sqrtz}d\fz^{-2}\\
\nonumber&&-\int_{\infty}^1\frac{2\tq^2\lagb(1-4\lagb)\fz}{\sqrtz}\frac{C_0\Nsh^{-2}\tom_1^2}{4\tq^2(1-4\lagb)}\ln\frac{1-\sqrtz}{1-\sqrt{1-4\lagb}}d\fz\\
\nonumber&&+\frac{4C_0\Nsh^{-2}\tom_1\tom_2\lagb}{\tq^2(1-4\lagb)}\left.(\ln f)\right|_{\fz=1}-\frac{C_0\Nsh^{-2}\tom_1^2\lagb^2}{1+\sqrt{1-4\lagb}}\int_{\infty}^1\frac{1}{1-\sqrtz}d\fz^{-2}\\
\nonumber&\simeq&\int_{\infty}^1\frac{2\Nsh^{-2}\tom_1^2\lagb}{\tq^2(1-4\lagb)}\Biggl[\frac{2C_0\Nsh^{-2}\tom_1^2\lagb}{\tq^2\left(1-\sqrt{1-4\lagb}\right)}+C_1+\frac{C_0\tq^2(1-4\lagb)}{16\left(1+\sqrt{1-4\lagb}\right)}\left(\ln\frac{16}{1-4\lagb}-2\ln f\right)\Biggr]d\ln f\\
\nonumber&&-\frac{C_0\Nsh^{-2}\tom_1^2\lagb^2}{\sqrt{1-4\lagb}\left(1+\sqrt{1-4\lagb}\right)}\int_{\infty}^1\frac{1-\sqrt{1-4\lagb}}{1-\sqrtz}d\fz^{-2}\\
\nonumber&&-\frac{C_0\Nsh^{-2}\tom_1^2\lagb^2}{1-4\lagb}\int_{\infty}^1\frac{1}{1-\sqrtz}d\fz^{-2}-\frac{C_0\Nsh^{-2}\tom_1^2\lagb}{4(1-4\lagb)}\left.(\ln f)\right|_{\fz=1}\\
\nonumber&&+\frac{4C_0\Nsh^{-2}\tom_1\tom_2\lagb}{\tq^2(1-4\lagb)}\left.(\ln f)\right|_{\fz=1}-\frac{C_0\Nsh^{-2}\tom_1^2\lagb^2}{1+\sqrt{1-4\lagb}}\int_{\infty}^1\frac{1}{1-\sqrtz}d\fz^{-2}\\
\nonumber&\simeq&\int_{\infty}^1\frac{2\Nsh^{-2}\tom_1^2\lagb}{\tq^2(1-4\lagb)}\Biggl[\frac{2C_0\Nsh^{-2}\tom_1^2\lagb}{\tq^2\left(1-\sqrt{1-4\lagb}\right)}+C_1+\frac{C_0\tq^2(1-4\lagb)}{16\left(1+\sqrt{1-4\lagb}\right)}\left(\ln\frac{16}{1-4\lagb}-2\ln f\right)\Biggr]d\ln f\\
\nonumber&&-\left[\frac{C_0\Nsh^{-2}\tom_1^2\lagb^2}{\sqrt{1-4\lagb}\left(1+\sqrt{1-4\lagb}\right)}+\frac{C_0\Nsh^{-2}\tom_1^2\lagb^2}{1-4\lagb}\right]\int_{\infty}^1\Biggl(\frac{1}{\sqrtz}\\
\nonumber&&+\frac{1}{1-\sqrtz}\Biggr)d\fz^{-2}-\frac{C_0\Nsh^{-2}\tom_1^2\lagb}{4(1-4\lagb)}\left.(\ln f)\right|_{\fz=1}+\frac{4C_0\Nsh^{-2}\tom_1\tom_2\lagb}{\tq^2(1-4\lagb)}\left.(\ln f)\right|_{\fz=1}\\
%\nonumber&\simeq&\int_{\infty}^1\frac{2\Nsh^{-2}\tom_1^2\lagb}{\tq^2(1-4\lagb)}\Biggl[\frac{2C_0\Nsh^{-2}\tom_1^2\lagb}{\tq^2\left(1-\sqrt{1-4\lagb}\right)}+C_1+\frac{C_0\tq^2(1-4\lagb)}{16\left(1+\sqrt{1-4\lagb}\right)}\left(\ln\frac{16}{1-4\lagb}-2\ln f\right)\Biggr]d\ln f\\
%\nonumber&&-\frac{1}{8\lagb}\left[\frac{C_0\Nsh^{-2}\tom_1^2\lagb^2}{\sqrt{1-4\lagb}\left(1+\sqrt{1-4\lagb}\right)}+\frac{C_0\Nsh^{-2}\tom_1^2\lagb^2}{1-4\lagb}\right]\left.\lncom\right|_{\fz=1}\\
%\nonumber&&-\frac{C_0\Nsh^{-2}\tom_1^2\lagb}{4(1-4\lagb)}\left.(\ln f)\right|_{\fz=1}+\frac{4C_0\Nsh^{-2}\tom_1\tom_2\lagb}{\tq^2(1-4\lagb)}\left.(\ln f)\right|_{\fz=1}\\
\nonumber&\simeq&\int_{\infty}^1\frac{2\Nsh^{-2}\tom_1^2\lagb}{\tq^2(1-4\lagb)}\Biggl[\frac{2C_0\Nsh^{-2}\tom_1^2\lagb}{\tq^2\left(1-\sqrt{1-4\lagb}\right)}+C_1+\frac{C_0\tq^2(1-4\lagb)}{16\left(1+\sqrt{1-4\lagb}\right)}\left(\ln\frac{16}{1-4\lagb}-2\ln f\right)\Biggr]d\ln f\\
\nonumber&&+\frac{1}{4\lagb}\left[\frac{C_0\Nsh^{-2}\tom_1^2\lagb^2}{\sqrt{1-4\lagb}\left(1+\sqrt{1-4\lagb}\right)}+\frac{C_0\Nsh^{-2}\tom_1^2\lagb^2}{1-4\lagb}\right]\left.(\ln f)\right|_{\fz=1}\\
\nonumber&&-\frac{C_0\Nsh^{-2}\tom_1^2\lagb}{4(1-4\lagb)}\left.(\ln f)\right|_{\fz=1}+\frac{4C_0\Nsh^{-2}\tom_1\tom_2\lagb}{\tq^2(1-4\lagb)}\left.(\ln f)\right|_{\fz=1}\\
\nonumber&\simeq&\Biggl[\frac{4C_0\Nsh^{-4}\tom_1^4\lagb^2}{\tq^4(1-4\lagb)\left(1-\sqrt{1-4\lagb}\right)}+\frac{2C_1\Nsh^{-2}\tom_1^2\lagb}{\tq^2(1-4\lagb)}+\frac{C_0\Nsh^{-2}\tom_1^2\lagb}{8\left(1+\sqrt{1-4\lagb}\right)}\ln\frac{16}{1-4\lagb}\\
&&+\frac{C_0\Nsh^{-2}\tom_1^2\lagb}{4\sqrt{1-4\lagb}\left(1+\sqrt{1-4\lagb}\right)}+\frac{4C_0\Nsh^{-2}\tom_1\tom_2\lagb}{\tq^2(1-4\lagb)}\Biggr]\left.(\ln f)\right|_{\fz=1}-\frac{C_0\Nsh^{-2}\tom_1^2\lagb}{8\left(1+\sqrt{1-4\lagb}\right)}\left.(\ln f)^2\right|_{\fz=1}
\end{eqnarray}
In Eq. \eqref{g2div}, these divergent terms are assembled according to Eq. \eqref{g2v}.

%comment20181225

%comment20181225

\end{document}